\documentclass[AMA,STIX2COL,Linenumbersoff]{MRM}
\articletype{Research Article}%

\usepackage{pifont}%
\usepackage[utf8]{inputenc} %
\usepackage[T1]{fontenc} %
\usepackage{tabularx}
\usepackage{nameref}
\usepackage{varioref}
\usepackage{hyperref}
\usepackage{cleveref}
\usepackage{graphicx}
\usepackage{float}
\usepackage{epstopdf}
\usepackage{multirow}
\usepackage{amssymb}
\usepackage{amsmath}
\usepackage{amsthm}
\usepackage{mathtools, cases}
\usepackage{chngcntr}
\usepackage{etoolbox}%
\usepackage{url}
\usepackage{subcaption}
\usepackage[font=scriptsize,labelfont=scriptsize]{caption}
\usepackage{color}
\usepackage[dvips]{epsfig}
\usepackage[export]{adjustbox}
\usepackage{bm}
\usepackage{tensor}
\usepackage{makecell}
\newcommand{\C}{\mathbb C}  %
\newcommand{\R}{\mathbb R}  %
\newcommand{\N}{\mathcal N} %
\newcommand{\Loss}{\mathcal L} %

\DeclarePairedDelimiterX{\norm}[1]{\lVert}{\rVert}{#1} 
\DeclarePairedDelimiterX{\abs}[1]{\lvert}{\rvert}{#1}
\DeclarePairedDelimiterX{\ip}[2]{\langle}{\rangle}{#1, #2} %
\DeclarePairedDelimiterX{\pfrac}[2]{(}{)}{\frac{#1}{#2}}
\DeclareMathOperator{\sign}{sign}
\DeclareMathOperator*{\argmin}{arg\,min\,} %

\DeclareMathOperator{\maximum}{\mathrm{max}\,}

\DeclareMathOperator{\suchthat}{\mathrm{s.t.}\,}
\DeclareMathOperator{\ST}{\mathrm{ST}}

\DeclareMathOperator{\RE}{Re}

\DeclareMathOperator{\diag}{\mathbf{diag}} %

\DeclareMathOperator{\Cov}{\mathrm{Cov}} %
\DeclareMathOperator{\E}{\mathbb{E}}     %

\newcommand{\cwise}[1]{\underline{#1}} %
\newcommand{\pwise}[1]{\overline{#1}} %

\newcommand{\bbb}{\bm{b}}
\newcommand{\bee}{\bm{e}}
\newcommand{\bg}{\bm{g}}

\newcommand{\bk}{\bm{k}}

\newcommand{\bs}{\bm{s}}

\newcommand{\bu}{\bm{u}}
\newcommand{\bv}{\bm{v}}

\newcommand{\x}{\bm{x}}
\newcommand{\y}{\bm{y}}
\newcommand{\z}{\bm{z}}

\newcommand{\bA}{\bm{A}}
\newcommand{\bB}{\bm{B}}
\newcommand{\bC}{\bm{C}}
\newcommand{\bD}{\bm{D}}

\newcommand{\bF}{\bm{F}}
\newcommand{\bG}{\bm{G}}

\newcommand{\bL}{\bm{L}}
\newcommand{\bM}{\bm{M}}

\newcommand{\bR}{\bm{R}}
\newcommand{\bS}{\bm{S}}

\newcommand{\bU}{\bm{U}}
\newcommand{\bV}{\bm{V}}

\newcommand{\bW}{\bm{W}}

\newcommand{\bZ}{\bm{Z}}

\newcommand{\btau}{\boldsymbol{\tau}}
\newcommand{\bmu}{\boldsymbol{\mu}}

\newcommand{\bxi}{\boldsymbol{\xi}}

\newcommand{\bvphi}{\boldsymbol{\varphi}}
\newcommand{\bsigma}{\boldsymbol{\sigma}}

\newcommand{\bSigma}{\bm{\Sigma}}
\newcommand{\bLambda}{\bm{\Lambda}}

\newcommand{\isqrtSigma}{\bSigma^{-1/2}}

\newcommand{\bOne}{\bm{1}}
\newcommand{\IDMAT}{\bm{I}}

\usepackage[algo2e,ruled,vlined]{algorithm2e}
\SetInd{0em}{2em}
\SetKwComment{Comment}{  \# }{}

\usepackage{array}
\newcolumntype{H}{>{\setbox0=\hbox\bgroup}c<{\egroup}@{}}
\usepackage{siunitx}
\let\cleardoublepage=\clearpage

\begin{document}
\title{Self-Supervised Noise Adaptive MRI Denoising via Repetition to Repetition (Rep2Rep) Learning}
\author[1,2]{Nikola Janju\v{s}evi\'{c}}{\orcid{0000-0002-8277-3175}}
\author[1,2]{Jingjia Chen}{\orcid{0000-0002-6023-1033}}
\author[1,2]{Luke Ginocchio}{}
\author[1,2]{Mary Bruno}{}
\author[1,2]{Yuhui Huang}{}
\author[3]{Yao Wang}{\orcid{0000-0003-3199-3802}}
\author[1,2]{Hersh Chandarana}{}
\author[1,2]{Li Feng}{\orcid{0000-0002-8692-7645}}
\authormark{\textsc{Janju\v{s}evi\'{c}} \textsc{et al.}}

\address[1]{\orgdiv{Bernard and Irene Schwartz Center for Biomedical Imaging, Department of Radiology}, \orgname{New York University Grossman School of Medicine}, \orgaddress{New York, \state{New York}, \country{USA}}}
\address[2]{\orgdiv{Center for Advanced Imaging Innovation and Research (CAI2R), Department of Radiology}, \orgname{New York University Grossman School of Medicine}, \orgaddress{New York, \state{New York}, \country{USA}}}
\address[3]{\orgdiv{Video Lab, Department of Electrical and Computer Engineering}, \orgname{New York University Tandon School of Engineering}, \orgaddress{Brooklyn, \state{New York}, \country{USA}}}
\corres{Nikola Janju\v{s}evi\'{c}. \email{nikola.janjusevic@nyulangone.org}}

\finfo{%
    This work was supported by the \fundingAgency{National Institute of Health}
    (\fundingNumber{R01EB030549}, \fundingNumber{R01EB031083},
    \fundingNumber{R21EB032917}, and \fundingNumber{P41EB017183}).
}

\abstract[Summary]{%
\section{Purpose} 
This work proposes a novel self-supervised noise-adaptive image denoising
framework, called Repetition to Repetition (Rep2Rep) learning, for low-field (<1T) MRI applications.

\section{Methods} 
Rep2Rep learning extends the Noise2Noise framework by training a neural network
on two repeated MRI acquisitions, using one repetition as input and another as
target, without requiring ground-truth data. It incorporates 
noise-adaptive training, enabling denoising generalization across
varying noise levels and flexible inference
with any number of repetitions. Performance was evaluated on both synthetic
noisy brain MRI and 0.55T prostate MRI data, and compared against supervised learning and
Monte Carlo Stein’s Unbiased Risk Estimator
(MC-SURE).

\section{Results} 
Rep2Rep learning outperforms MC-SURE on both synthetic and 0.55T
MRI datasets. On synthetic brain data, it achieved denoising quality comparable
to supervised learning and surpassed MC-SURE, particularly in preserving
structural details and reducing residual noise. On the 0.55T prostate MRI
dataset, a reader study showed radiologists preferred Rep2Rep-denoised
2-average images over 8-average noisy images. Rep2Rep demonstrated 
robustness to noise-level discrepancies between training and inference,
supporting its practical implementation.

\section{Conclusion} 
Rep2Rep learning offers an effective self-supervised denoising for
low-field MRI by leveraging routinely acquired multi-repetition data. Its
noise-adaptivity enables generalization to different SNR regimes
without clean reference images. This makes Rep2Rep learning a
promising tool for improving image quality and scan efficiency in low-field
MRI.
}
\wordcount{4991}

\jnlcitation{%
\cname{%
\author{N. Janju\v{s}evi\'{c}}, 
\author{J. Chen}, 
\author{L. Ginocchio},
\author{M. Bruno},
\author{Y. Huang},
\author{Y. Wang},
\author{H. Chandarana}, and
\author{Li Feng}
(\cyear{2025}), 
\ctitle{Self-Supervised Noise Adaptive MRI Denoising via Repetition to Repetition (Rep2Rep) Learning}, 
\cjournal{Magn. Reson. Med.}, 
\cvol{xxxx;xx:xx-xx}.
}}

\maketitle

\section{Introduction}\label{sec:intro}
Low-field MRI, particularly at 0.55T or lower field strengths, has gained increasing attention
in both research and clinical applications due to its unique advantages~\cite{Shetty2023, Hori2021}.
Compared to standard clinical MRI systems operated at 1.5T-3T, low-field MRI
offers reduced system costs, lower operational expenses, and improved field
homogeneity, which can facilitate new applications that are traditionally
challenging at higher field~\cite{Tian2023}. However, one of the main challenges associated
with low-field MRI is the significant reduction in signal-to-noise ratio (SNR),
which limits its widespread clinical adoption~\cite{Arnold2022}.

A simple way to improve SNR in low-field MRI is to increase the number of
signal averages during data acquisition, which involves repeating the MRI
scan multiple times (referred to as repetitions hereafter) and averaging the
results to reduce noise at the cost of prolonged scan times. However, this
approach has several limitations, as long scan time increases patient
discomfort, operational costs, and the risk of motion artifacts, which may lead
to degraded image quality~\cite{Shetty2023,Arnold2022}. Therefore, alternative strategies are highly
preferred to enhance SNR in low-field MRI without compromising scan efficiency.

A promising solution to address this challenge is deep learning-based
denoising, which aims to recover high-SNR images from noisy acquisitions.
However, a major challenge in applying standard deep learning methods for MR images
acquired at low field is the lack of high-SNR ground truth images for
supervised training~\cite{TIAN2022}. This necessitates self-supervised learning strategies,
which can learn denoising directly from noisy images without
requiring clean references. Several self-supervised learning approaches have
been proposed for MRI denoising. One widely used method is Stein’s Unbiased
Risk Estimator (SURE)~\cite{MCSURE2008, SUREdn}, which enables self-supervised learning by approximating
the mean squared error using statistical properties of noise. This approach has
been demonstrated by Pfaff et al for low-field MRI applications~\cite{Pfaff2023, Pfaff2024}. While
effective, SURE-based methods require knowledge of the image noise-level (see Section \ref{sec-theory-training}),
rendering it sensitive to noise-level estimation error~\cite{tachella2025unsureselfsupervisedlearningunknown}. Moreover, SURE assumes
that noise is spatially uncorrelated, which is not always valid in MR images
due to different reasons, such as the use of parallel imaging
techniques that introduce spatial noise correlations~\cite{AjaFernndez2010}. These limitations
restrict the ultimate performance of SURE-based denoising in low-field MRI.

An alternative self-supervised denoising strategy is the Noise2Noise framework~\cite{Noise2Noise_Lehtinen2018},
which eliminates the need for noise level estimation by training a model using
pairs of independently noisy images of the same underlying signal. This
approach is well-suited for low-field MRI, where acquiring multiple repetitions of
the same scan with independent noise, for purposes of noise reduction via repetition averaging, is common practice~\cite{Arnold2022}.
However, a critical
limitation of standard Noise2Noise learning is the discrepancy between training
and inference. During training, one noisy repetition serves as the input while
the other acts as the target, whereas at inference, both repetitions are
typically combined to enhance the baseline SNR. This mismatch can lead to
suboptimal generalization during deployment.

To address these challenges, we propose a novel self-supervised MRI denoising
framework, called Repetition to Repetition (Rep2Rep) Learning, for low-field MRI
applications. Rep2Rep extends the Noise2Noise paradigm by training the model on
repeated acquisitions while incorporating a noise-adaptive training strategy,
allowing the network to generalize across varying noise levels.
Moreover, we demonstrate that Rep2Rep learning is inherently compatible with
parallel imaging, such as GRAPPA (Generalized Autocalibrating Partially
Parallel Acquisitions)~\cite{GRAPPA_Griswold2002}, and it can effectively handle spatially correlated
noise. The performance of Rep2Rep learning
is validated through both synthetic noisy brain MRI and low-field
(0.55T) prostate MRI for evaluation.

The remainder of this paper is structured as follows: Section \ref{sec-theory}
describes the theoretical background on multicoil MRI noise modeling,
self-supervised denoising strategies, and noise-adaptive deep denoising. Section \ref{sec-methods} presents the
proposed Rep2Rep learning framework, including data preprocessing, training, and inference strategy. 
Section \ref{sec-results} provides
quantitative and qualitative evaluations, comparing Rep2Rep against existing
methods on both synthetic and real low-field MRI datasets. Finally, Section
\ref{sec-discussion} discusses the implications and future directions of this
work. 

\section{Theory}\label{sec-theory}
\begin{table}[tb]
\centering
\caption{Notation. \label{tab:notation}}
\vspace{0.5em}
\resizebox{\linewidth}{!}{%
\begin{tabular}{|l|l|} \hline
\multirow{2}{*}{$\x \in \C^{NC}$} & a vector valued image with $N=N_1\times N_2$ pixels, 
\\ & and vectorized channels, $\x =[ \x_1^T,\, \cdots,\, \x_C^T ]^T$. \\
\hline
$\x_c \in \C^N$ & the $c$-th channel/coil of $\x$. \\
\hline
$\x[n] \in \C^C$ & the $n$-th pixel of $\x$, $n \in [1, \, N]$. \\
\hline
$\x_c[n] \in \C$ & the $n$-th pixel of the $c$-th channel of $\x$. \\
\hline
$\bu \circ \bv \in \C^N$ & the element-wise product of two
vectors. \\
\hline
$\bU \in \C^{Q\times N}$ & a $Q \times N$ matrix with elements $\bU_{ij} \in \C$. \\
\hline
$\bU_{i:} \in \C^N,~\bU_{:j} \in \C^Q$ & the $i$-th row, $j$-th column of matrix $\bU$. \\
\hline
\multirow{2}{*}{$\bU \otimes \bV \in \C^{QM \times NC}$} & Kronecker product of $\bU \in
\C^{Q\times N}$ and $\bV \in \C^{M\times C}$, \\
& i.e. the block matrix with $\bV$s scaled by $\bU_{ij} ~ \forall ~ i,j$.\\
\hline
\multirow{2}{*}{$\y = \cwise{\bU}\x \equiv (\bU \otimes \IDMAT_C)\x$} & the matrix $\bU$ applied channel-wise, \\
& i.e. $\y_c = \bU\x_c \in \C^Q,~\forall~1 \leq c \leq C$. \\
\hline
\multirow{2}{*}{$\y = \pwise{\bV}\x \equiv (\IDMAT_N \otimes \bV)\x$} & the matrix $\bV$ applied pixel-wise,\\
& i.e. $\y[i] = \bV\x[i] \in \C^M,~\forall~1 \leq i \leq N$. \\
\hline
\end{tabular}}
\end{table}
\subsection{Multicoil MRI Acquisition and Reconstruction}\label{sec-theory-mri}
Modern MRI scanners typically use multicoil receiver arrays for data
acquisition. Each coil captures complex-valued signals from the region of
interest, which are contaminated by thermal noise originating from both the
body of patients and electronics of the scanner. Furthermore, the noise may be
correlated between coil elements due to their proximity and shared electrical
circuitry~\cite{CrdenasBlanco2008}. In MRI, we are interested in obtaining an 
MR image ${\x \in \C^N}$ with $N$ pixels which is free from scanner noise.
We may mathematically describe this acquisition process as,
\begin{equation} 
    \bk_c = \bF(\bs_c \circ \x) + \bxi_c, \quad \forall ~ c=1,\, \dots,\, C,
\end{equation}
where ${\bk \in \C^{NC}}$ is the acquired $C$-coil k-space signal, $\bF$ is the
N-dimensional 2D DFT matrix, ${\bs_c \in \C^N}$ is the $c$-th coil's image-domain
sensitivity profile, and ${\bxi[n] \sim \N(0, \bSigma)}$ is complex valued
coil-noise with covariance matrix ${\bSigma \in \C^{C\times C}}$ identical in
each pixel location ${n \in [1,N]}$.
We write this compactly in the notation of Table \ref{tab:notation} as,
\begin{equation} 
    \bk = \cwise{\bF}\bS \x + \bxi, \quad \bxi \sim \N(0, \pwise{\bSigma}),
\end{equation}
where 
$\bS = \begin{bmatrix} \diag(\bs_1) & \cdots & \diag(\bs_C) \end{bmatrix}^T$ is
the coil sensitivity map operator,
$\cwise{\bF}$ represents the $N$-dimensional 2D-DFT matrix repeated channel-wise, 
and $\pwise{\bSigma}$ represents the noise covariance matrix repeated pixel-wise (see Supporting Information for more detailed discussion on notation).

Coil sensitivity maps may be readily estimated from fully-sampled low-frequency k-space
data, often referred to as the Auto-Calibration Signal (ACS), via the
ESPIRiT algorithm \cite{espirit}. 
We obtain the coil-combined noisy image as,
\begin{equation}\label{eq-full_reco}
    \begin{aligned}
    \y &= \bS^H\cwise{\bF}^H\bk \\
        &= \bS^H\cwise{\bF}^H(\cwise{\bF}\bS\x + \bxi) \\
        &= \x + \bxi^\prime,
    \end{aligned}
\end{equation}
where $\bxi^\prime \sim \N(0,\, \bS^H\pwise{\bSigma}\bS)$ and this noise
covariance is a diagonal matrix. The square-root of this diagonal is referred to as the
coil-combined image noise-level in this work, which is described by,
\begin{equation}\label{eq-noise_fully}
\bsigma[n] = \sqrt{\bs[n]^H\bSigma\bs[n]}, \quad \forall ~ n = 1, \, \dots, \, N,
\end{equation}
or, written in the notation of Table \ref{tab:notation} (see Supporting Information), ${\bsigma =
\sqrt{\bS^H\pwise{\bSigma}\bs} \in \R^N_+}$, where $\sqrt{\cdot}$ is taken elementwise.
Note that this noise-level may be spatially varying due to differing noise
strengths in each coil and/or noise correlations between coils~\cite{aja-fernandez_statistical_2011}.

To reduce motion sensitivity within each repetition, parallel imaging can be used to accelerate the acquisition, typically with an acceleration rate of 2-3. This can be described as,
\begin{equation}\label{eq-obs_masked}
    \bk = \cwise{\bM}(\cwise{\bF}\bS \x + \bxi), \quad \bxi \sim \N(0, \pwise{\bSigma}),
\end{equation}
where ${\bM \in \{0,1\}^{N\times N}}$ is the k-space mask operator describing
acquired sample locations. Low-acceleration rates are adequately
handled by linear reconstruction methods such as GRAPPA, which uses a data
specific interpolation operator $\bG^H$ to create a ``complete'' k-space,
$\bk^\prime = \bG^H\bk$. As with coil-sensitivities, this operator is
readily computed from the ACS. The corresponding coil-combined
image-domain reconstruction is described as follows,
\begin{equation}\label{eq-grappa_reco}
    \begin{aligned}
    \y &= \bS^H\cwise{\bF}^H\bk^\prime \\
    &= \bS^H\cwise{\bF}^H\bG^H\cwise{\bM}(\cwise{\bF}\bS\x + \bxi) \\
    &= \x_\mathrm{reco} + \bxi_\mathrm{reco},
    \end{aligned}
\end{equation}
where $\x_\mathrm{reco}$ is the GRAPPA reconstructed noise-free data and
${\bxi_\mathrm{reco} \sim \N(0,\, \bS^H\cwise{\bF}^H\bG^H(\bM \otimes
\bSigma)\bG\cwise{\bF}\bS)}$ is the coil-combined image-domain noise. It can be
shown that this noise covariance matrix is not diagonal any more, and spatial
correlations exist between noise-pixels depending on the sampling pattern
encoded by $\bM$~\cite{aja-fernandez_statistical_2011}~(see Supporting Information for more details). 
When $\bM$ encodes a regular Cartesian k-space undersampling without inclusion
of the ACS region, the image domain
coil-combined noise-level is given by,
\begin{equation}\label{eq-grappa_noise}
    \bsigma = \sqrt{\tfrac{1}{A} \cdot \bS^H\cwise{\bF}^H\bG^H\pwise{\bSigma}\bG\cwise{\bF}\bs},
\end{equation}
where $A$ is the acceleration factor. 

If the ACS is included in reconstruction, 
which is often desired to increase the resulting SNR, 
both local and non-local spatial noise correlations will be introduced~\cite{aja-fernandez_statistical_2011} (see Supporting Information). 
The image domain coil-combined noise-level of the GRAPPA+ACS reconstruction is given as,
\begin{equation}\label{eq-grappa_noise_acs}
    \bsigma = \sqrt{\tfrac{(1-p)}{A} \cdot \bS^H\cwise{\bF}^H\bG^H\pwise{\bSigma}\bG\cwise{\bF}\bs 
    + p \cdot \bS^H\pwise{\bSigma}\bs },
\end{equation}
where $p$ is the fraction of the ACS region in k-space.

\subsection{From Supervised Denoising to Self-Supervised Denoising}\label{sec-theory-training}
A parameterized denoiser $f$ can be
optimized by minimizing expected mean-squared-error (MSE) over a distribution
of noisy ($\y$) and ground-truth ($\x$) image pairs,
\begin{equation} \label{eq-loss_mse}
    \Loss_{\mathrm{MSE}} = \E[ \norm{\x - f(\y)}_2^2 ].
\end{equation}
However, collecting ground-truth data may require a
large number of repetitions in low-SNR MRI, which can be challenging due to
cost, time, and the resulting motion artifacts from extended scan time.
To overcome the need for ground-truth training data, numerous methods have been developed which exploit noise-distribution knowledge~\cite{MCSURE2008,SUREdn,tachella2025unsureselfsupervisedlearningunknown}, network architectures~\cite{krull_noise2void_2019}, or noisy data-partitioning~\cite{Noise2Noise_Lehtinen2018,yaman_self-supervised_2020,batson2019noise2self}. In this work, we look at two representative methods for self-supervised training: MC-SURE~\cite{MCSURE2008} and Noise2Noise~\cite{Noise2Noise_Lehtinen2018}.

Stein's Unbiased Risk Estimator
(SURE) offers access to the MSE loss function without ground-truth training given that our
data is contaminated with uncorrelated Gaussian noise. Consider $\y = \x +
\bxi^\prime$ with $\bxi^\prime \sim \N(0, \, \diag(\bsigma^2))$. The SURE loss
rewrites the supervised MSE loss as,
\begin{align} \label{eq-SURE}
    &~~\Loss_{\mathrm{MSE}} = \E[\norm{\x - f(\y)}_2^2] \\
    &= \E[\norm{\x - \y + \y - f(\y)}_2^2] \\
    &= \E[ \norm{\y - f(\y)}_2^2 - \bOne^T\bsigma^2 + 2\ip{\bxi^\prime}{f(\y)}] \label{eq:stein1}\\
    &= \E[ \norm{\y - f(\y)}_2^2 - \bOne^T\bsigma^2 + 2\nabla_{\y} \cdot (\bsigma^2 \circ f(\y))] \label{eq:stein2}\\
    &\coloneqq \Loss_{\mathrm{SURE}} 
\end{align} 
where $\bOne$ is the vector of all 1's and we've made use of the zero-mean and
signal-independent property of the noise, as well as Stein's lemma, to equate
the SURE loss to the MSE loss.
Note that this loss assumes a {\it spatially uncorrelated noise} from Equation \eqref{eq:stein1} to \eqref{eq:stein2} and exact
noise-level information. 

A Monte-Carlo method is used to estimate the SURE loss divergence
term via a finite difference. For complex-valued noisy observations we
draw $\bbb \in \C^N \sim \N(0, I)$, 
\begin{equation} \label{eq-cplx_mc_divergence}
    \nabla_{\y} \cdot f(\y) \approx \RE\left\{ \bbb^H \pfrac*{f(\y + h\bbb) - f(\y)}{h} \right\},
\end{equation}
where $h$ is manually set somewhere in [\num{e-5}, \num{e-3}]. For the rest of
the manuscript we refer to this approach as MC-SURE. 

When multiple repetitions of image data are available, the Noise2Noise model~\cite{Noise2Noise_Lehtinen2018} can be utilized to derive an MSE-equivalent loss function.
Consider two noisy observations of the same
underlying signal, ${\y = \x + \bxi}$ and ${\bv = \x + \bvphi}$, where $\bxi$
and $\bvphi$ are zero-mean, possibly spatially varying, noise vectors which are
independent of the signal and each other. Then,
\begin{align}
    &~~\Loss_{\mathrm{MSE}} = \E[\norm{\x - f(\y)}_2^2] \\
    &= \E[\norm{\x - \bv + \bv - f(\y)}_2^2] \\
    &= \E[2\ip{\x - \bv}{\bv - f(\y)} + \norm{\bv - f(\y)}_2^2] + \mathrm{const.}\\
    &= \E[\norm{\bv - f(\y)}_2^2 - 2\ip{\bvphi}{\bv - f(\y)} ] + \mathrm{const.}\\
    &= \E[\norm{\bv - f(\y)}_2^2 + \mathrm{const.}\\
    &\coloneqq \Loss_{\mathrm{N2N}} + \mathrm{const.}
\end{align}
Here, the constant terms with respect to the neural network parameters are
omitted, and the zero-mean and independence properties of the noise allow the
cross-terms to cancel, giving us access ot the supervised MSE loss in
expectation. These conditions are easily satisfied in complex-valued MRI data
acquired with multiple repetitions.

When using MRI
repetition data to train a denoiser, we may wish to average all available
repetitions at inference to improve SNR. This changes the noise-level
distribution relative to training which can lead to
degradation of denoising performance if not properly addressed~\cite{Mohan2020Robust,
janjusevicCDLNet2022}.

\subsection{Self-Supervised Noise-Adaptive Denoising}\label{sec-theory-ada}
Consider a classical sparsity-prior-based denoiser $f$ which seeks a sparse
representation $\z$ with respect to some given dictionary $\bD$,
\begin{equation}
    \begin{gathered}
        f(\y) = \bD\z^\ast \\
        \z^\ast = \argmin_{\z} \norm{\z}_1 ~ \suchthat ~ \norm{\y - \bD\z}_2^2 \leq \epsilon^2  \\
        = \argmin_{\z} \lambda\norm{\z}_1 + \tfrac{1}{2}\norm{\y - \bD\z}_2^2,
    \end{gathered}
\end{equation}
where $\norm{\cdot}_1$ imposes a sparsity constraint, and $\lambda >0$ is a
Lagrange multiplier term  related to the observation noise-level
$\bsigma$. An algorithm for solving this optimization is the Iterative
Soft-Thresholding algorithm (ISTA)~\cite{Beck2009}, which iterates $\z^{(k+1)} =
\ST_{\eta\lambda}(\z^{(k)} - \eta \bD^H (\bD \z^{(k)} - \y))$ until
convergence, where $\eta > 0$ is a step-size parameter, $\ST_\tau(\z) =
\sign(\z)\max(0,\, \abs{\z}-\tau)$ is the elementwise soft-thresholding
operator, and $\Theta = \{\bD, \, \lambda\}$ represent the denoiser parameters. 
The choice of $\lambda$ depends on the noise-level of the input
$\y$, making this classical denoiser noise-adaptive. 

We extended this ISTA-based denoiser by formulating the Convolutional Dictionary Learning
Network (CDLNet)~\cite{janjusevicCDLNet2022} deep learning based denoiser, which we demonstrated on natural image denoising. 
The CDLNet architecture is given as,
\begin{equation}
    \begin{gathered}
        f(\y) = \bD\z^{(K)}, \quad \z^{(0)} = \bm{0} \\
        \z^{(k+1)} = \ST_{\btau^{(k)}}(\z^{(k)} - {\bA^{(k)}}^H (\bB^{(k)} \z^{(k)} - \y))
    \end{gathered}
\end{equation}
where $\bA$ and $\bB$ are convolutional dictionaries and $\btau$ are subband
dependent thresholds. CDLNet achieves noise-adaptivity through an affine parameterization of the thresholds,
\begin{equation}\label{eq-cdl_noise_ada}
    \btau^{(k)} = \btau_0^{(k)} + \bsigma \btau_1^{(k)},
\end{equation}
where a pre-computed noise-level $\bsigma$ is used as side-information to
update the threshold weights of the network. The parameters of the
network are given by
${\Theta = \{\bD,\, \{ {\bA^{(k)}},\, \bB^{(k)},\, \btau_0^{(k)},\, \btau_1^{(k)} \}_{k=0}^{(K-1)} \}}$.

Unlike standard neural networks that often lack explicit mechanisms for noise
adaptivity or rely on concatenating a noise-level map as an additional input~\cite{FFDNet},
CDLNet incorporates noise information directly into its layerwise thresholding.
Prior studies~\cite{Mohan2020Robust,janjusevicCDLNet2022,janjusevicGDLNet2022,janjusevicGroupCDL2025}
have shown that such implicit noise-adaptive result in suboptimal performance
within the training noise range and catastrophic failure outside it, whereas CDLNet extrapolates near-perfectly~\cite{janjusevicCDLNet2022}.

\begin{figure*}[tbh]
\centerline{\includegraphics[width=\textwidth]{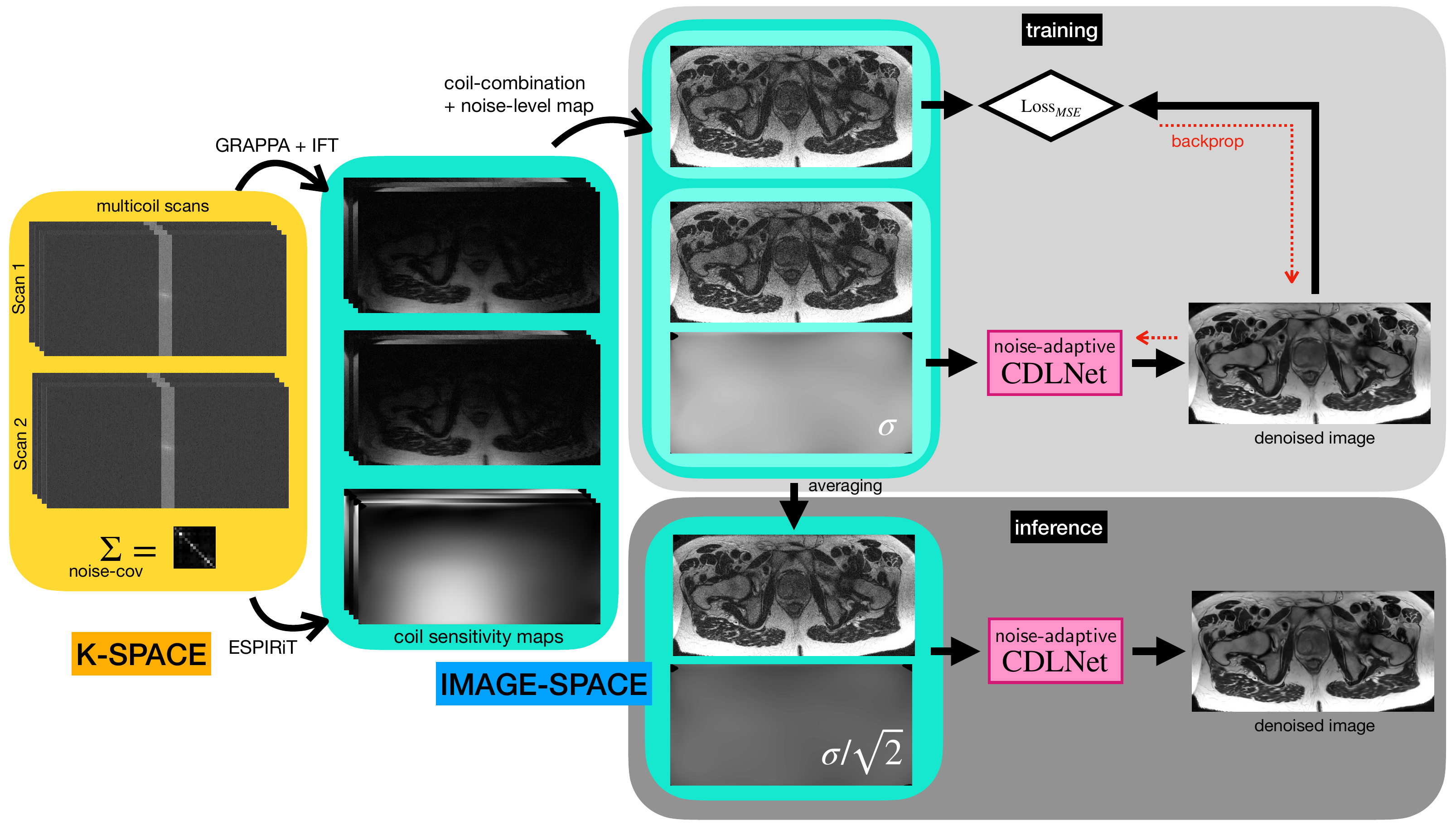}}
\caption{%
    The Rep2Rep learning framework with noise-adaptive inference. We start with
    two repetitions of multicoil k-space (scan 1 and scan 2), which we perform
    noise-level estimation to obtain a noise coil covariance matrix $\bSigma$.
    We then perform reconstruction (with GRAPPA if necessary) and coil
    sensitivity map estimation using ESPIRiT. The multicoil repetition images
    are then coil-combined and labeled with an associated noise-level map
    (using Equations~\eqref{eq-noise_fully}~or~\eqref{eq-grappa_noise_acs}).
    During Rep2Rep training, one repetition image serves as input and one as
    target, with noise-level map used as side-information to train a CDLNet
    denoiser. At inference, repetitions are averaged to optimally improve
    denoising results.
\label{fig:snaper}}
\end{figure*}
\section{Methods}\label{sec-methods}
\subsection{The Rep2Rep learning Framework}\label{sec-methods-r2r}
We propose a self-supervised MRI denoising framework based on Repetition to
Repetition (Rep2Rep) learning to enable noise-adaptive MRI repetition denoising.
Given a dataset of low-SNR multicoil k-space samples, including at least two
repetitions, our goal is to train a noise-level-adaptive denoiser that
generalizes across varying acquisition conditions. 
Given repeittion data, we first estimate the kspace noise-covariance matrix
(see Supporting Informaiton) and then reconstruct images and perform coil
combination, using Equations \eqref{eq-noise_fully}, \eqref{eq-grappa_noise},
or \eqref{eq-grappa_noise_acs} to label our data with its associated
image-domain noise level $(\bsigma)$.
During training, a single repetition scan $\y_{(1)}$ and
its noise-level $\bsigma$ are used as input to a CDLNet denoiser, while the other
repetition $\y_{(2)}$ serves as the target. The training loss is defined as,
\begin{equation} \label{eq-loss_rep2rep}
    \Loss_{\mathrm{R2R}} = \norm{\y_{(2)} - f(\y_{(1)}; \, \bsigma)}_2^2
\end{equation}
Since MRI noise is zero mean and independent between repetitions, the Rep2Rep
loss adheres to the Noise2Noise paradigm, enabling the training of a
noise-adaptive CDLNet denoiser. At inference, any number of repetitions $R$ may
be averaged and denoised in a single step,
\begin{equation}\label{eq-rep2rep_inference}
    \hat{\x} = f \left(\sum_{r=1}^R \y_{(r)}; \frac{\bsigma}{\sqrt{R}} \right).
\end{equation}
The proposed Rep2Rep learning framework for training and inference is
summarized in Figure \ref{fig:snaper}.

\subsection{Post-Processing}\label{sec-methods-postproc}
It is well known that supervised and self-supervised denoiser training often
exhibit a tendency to over-smooth image details. This effect is particularly
pronounced in MSE-based training, where the denoiser converges to a minimum-MSE
solution by effectively averaging all plausible reconstructions that are
consistent with the observed noisy image. To mitigate this over-smoothing
effect, we apply a two-step post-processing strategy: first, a sharpening filter is used to enhance
fine image details by combining the denoised output with the original noisy
input,
\begin{equation}\label{eq-sharpen} 
    \hat{\x}_s = (1-a)\hat{\x} + a(\bm{h} \ast \hat{\x}),
\end{equation}
where $a$ (\%) controls the sharpening intensity. Next, we incorporate a
dithering process to generate subtle perturbations to mitigate unnatural
smoothness and restore fine textures in the denoised images, via,
\begin{align}\label{eq-dither}
    \hat{\x}_d &= (1-b\sqrt{N})\hat{\x}_s + b\sqrt{N} \y, %
\end{align}
where $b$ (\%) determines the dithering strength, and $b< 1 / \sqrt{N}$.

\subsection{Implementation}
The Rep2Rep learning framework was implemented in the Julia programming language with
Lux.jl~\cite{pal2023lux}. Complex-valued CDLNet models (of the same architecture as \cite{janjusevicCDLNet2022}) were used 
for synthetic-noise and real-noise experiments. Models were trained on image patches of $128\times 128$ 
and batch-size of 8 for 500k gradient steps. The initial learning rate was set to $\num{5e-4}$ using the Adam optimizer. 
Further training details can be found in \cite{janjusevicCDLNet2022}. All models were trained with an NVIDIA A100 GPU.

\subsection{Evaluation on Brain Datasets with Synthetic Noise}\label{sec-methods-brain}
The performance of Rep2Rep learning was first quantitatively evaluated on
T2-weighted brain datasets from the fastMRI database~\cite{fastmri}, where low-SNR multicoil
MRI images were generated with synthetic noise to facilitate a comparative
analysis of different training and inference strategies for denoising. A total
of 281 multi-slice datasets were used. For each dataset, we first estimated
coil-sensitivity maps, combined the fully sampled multicoil image, and
centered-cropped each slice to $320\times 320$ pixels to create a high-quality
single-coil complex image, which serves as the ground truth reference for
quantitative evaluations. 

We then generated noisy fully sampled repeittion data a single k-space
noise-covariance matrix for each volume (see Supporting Information).
Along with the fully sampled noisy images, we generated GRAPPA-reconstructed
noisy images with an undersampling factor of 2 and 24 central k-space lines as
the ACS data. Two GRAPPA-reconstructed images were generated: one including the
ACS data in the final image and the other excluding it. 

A total of 253 datasets were used for network training and
validation, while the remaining 28 datasets were used for evaluation. Each
dataset was labeled with both its exact noise-level information and an
imperfect noise-level estimate obtained by multicoil image-domain filtering to
obtain a coil-covariance estimate (see Supporting Information).
Denoisers were trained using this imperfect noise-level information unless otherwise specified.

We performed quantitative evaluation using standard image quality metrics: Normalized Root Mean Square Error
(NRMSE), expressed as a percentage, and Structural Similarity Index Measure
(SSIM) with a range of $[-1,1]$, where 1 indicates perfect similarity. In
addition, we measured how well the residual between the input noisy image $\y$
and the denoised image $\hat{\x}$ matches the known noise distribution by computing
the normalized residual variance (NRV),
\begin{equation}
    \frac{1}{N-1}\sum_{n=1}^N \abs*{\frac{\y[n] - \hat{\x}[n]}{\bsigma[n]}}^2.
\end{equation}
Here, a perfectly denoised image would yield NRV = 1, while suboptimal
denoising can result in values above or below 1. 
Competing methods were compared for statistical
significance using a Student's-t distribution test.

Rep2Rep learning was evaluated using three different training and inference
strategies: (1) pre-denoising repetition averaging with noise-adaptive
denoising (Pre-Avg+Ada); (2) post-denoising
repetition averaging (Post-Avg); (3) pre-denoising repetition
averaging without noise-adaptive denoising (Pre-Avg).
These strategies are defined as follows:
\begin{align}
    \hat{\x}_\mathrm{pre-avg+ada} &= f\left(\tfrac{1}{R}\sum_{r=1}^R \y_{(r)}; \, \tfrac{1}{\sqrt{R}}\bsigma \right), \label{eq-pre_ada}\\
    \hat{\x}_\mathrm{post-avg} &= \tfrac{1}{R}\sum_{r=1}^R f\left(\y_{(r)}; \, \bsigma \right), \label{eq-post} \\
    \hat{\x}_\mathrm{pre-avg} &= f\left(\tfrac{1}{R}\sum_{r=1}^R \y_{(r)}\right). \label{eq-pre}
\end{align}
Unless otherwise specified, the Pre-Avg+Ada strategy was used throughout the
manuscript in addition to this experiment.

\subsection{Evaluation on 0.55T Prostate Imaging}\label{sec-methods-prostate}
In this study, we evaluated the performance of Rep2Rep learning on real-world
T2-weighted (T2w) prostate images acquired at low field (0.55T). A total of 73
male subjects (mean age $= 68.56 \pm 12.52$) were recruited, with all
participants providing written informed consent prior to the MRI scans. Imaging
was performed on a clinical MRI scanner ramped down to 0.55T (MAGNETOM Aera,
Siemens Healthineers, Erlangen, Germany) using a multi-slice 2D fast spin-echo
(FSE) sequence. The imaging parameters were as follows: FOV = $360\times 180$ mm$^2$
mm$^2$, matrix size = $512\times 256$, in-plane spatial resolution = $0.7 \times 0.7$ mm$^2$,
slice thickness = 3 mm, and number of slices = 30. Each acquisition employed
2-fold GRAPPA acceleration with 24 ACS lines at the center of k-space.
Eight repetitions were acquired per dataset with the same undersampling mask,
with a total scan time of 14.48 minutes. 

53 cases were used for training and validation, while the remaining 20 cases
were used for evaluation. To process the data, we first estimated
coil-sensitivity maps from the ACS data, followed by noise covariance
estimation from a reference noise scan. The k-space data was then whitened (see
Supporting Information) and GRAPPA reconstructed. Unless otherwise specified,
the ACS region was included in the final reconstruction. Finally, we computed
the image-domain noise level according to Equation \eqref{eq-grappa_noise_acs}.
CDLNet denoisers were trained with a 1-repetition to 1-repetition Rep2Rep
learning scheme and all inferences were performed on 2-average repetition data
reconstructed {\it with the ACS region}, unless otherwise specified.

\subsection{Experimental Setup}
\noindent\textbf{Experiment 1}: 
{\it Comparison of Supervised Learning, MC-SURE, and Rep2Rep Learning in Fully Sampled Brain MRI with Synthetic Noise}.
We trained 3 CDLNet denoisers on the fully-sampled synthetic noise
brain dataset: one with a supervised MSE loss with 2-average noisy input data,
one with the MC-SURE loss with 2-average noisy input data, and one with a
Rep2Rep loss in a 1-repetition to 1-repetition scheme. As such, the
MC-SURE-trained and Rep2Rep trained models made use of the same amount of
training data. 

\noindent\textbf{Experiment 2}: 
{\it Comparison of Supervised Learning, MC-SURE, and Rep2Rep in 2× Accelerated GRAPPA-Reconstructed Brain MRI}.
We trained 3 CDLNet denoisers
on GRAPPA reconstructed synthetic noise brain datasets: one with a supervised
MSE loss with 2-average noisy input data (on GRAPPA+ACS data), one with the
MC-SURE loss with 2-average noisy input data (on GRAPPA without ACS data), and
one with a Rep2Rep loss in a 1-repetition to 1-repetition scheme (on GRAPPA+ACS
data). The discrepancy in training data between MC-SURE and other methods
(exclusion of the ACS) is due to the local noise-correlations introduced by
GRAPPA+ACS reconstruction, and is investigated further in the following experiment. 

\noindent\textbf{Experiment 3}: 
{\it Assessing the Impact of ACS Data and Noise Estimation on MC-SURE Denoising}
In addition to the Rep2Rep-trained and MC-SURE-trained models from the previous
experiment, trained on GRAPPA reconstructed data, we trained two other
denoisers with an MC-SURE loss: (1) with the ACS region included in the
training data; (2) with the ACS region excluded but perfect noise-level
information used during training. All other denoisers made use of estimated
noise-level information, and all denoisers used estimated noise-level
information at inference. Inferences were performed on 2-averaged GRAPPA+ACS
reconstructed data.

\noindent\textbf{Experiment 4}: 
{\it Comparison of Different Rep2Rep Training and Inference Strategies}
In addition to the Rep2Rep-trained fully-sampled data CDLNet denoiser used in
previous experiments, we trained a Rep2Rep-based CDLNet denoiser without
noise-adaptive thresholding (see Section \ref{sec-theory-ada}).

\noindent\textbf{Experiment 5}: 
{\it Evaluation of Post-Processing Settings in 0.55T Prostate MRI}
We trained a Rep2Rep-based denoiser in a 1-repetition to 1-repetition scheme on
the 0.55T T2w Prostate dataset. On the training set, we performed inference on
2-averaged data and post-processed the denoised images with various levels of
sharpening and dithering post-processing following the methods detailed in Section \ref{sec-methods-postproc}.

\noindent\textbf{Experiment 6}: 
{\it Comparison of Rep2Rep Learning and MC-SURE in 0.55T Prostate MRI}
This experiment followed the experimental setup of Experiment 3, now with denoisers trained on the 0.55T Prostate
dataset, to verify the validity of our synthetic-noise experiments on real
noisy data. Note that perfect noise-level information was not available.

\noindent\textbf{Experiment 7}: 
{\it Assessing the Impact of Repetition Number on Rep2Rep Training in 0.55T Prostate MRI}
In addition to the Rep2Rep-based denoiser trained in the previous 0.55T
Prostate data experiments (trained in a 1-repetition to 1-repetition scheme),
we trained two additional Rep2Rep-based denoisers in a 2-average to 2-average
scheme and 4-average to 4-average scheme. The effect of these different schemes
is increasing the SNR of input and target images during training. Inference was
performed on 2-average data.

\noindent\textbf{Experiment 8}: 
{\it Qualitative Evaluation of Denoising in 0.55T Prostate MRI}
Qualitative evaluation of the noisy and denoised images was performed through a
reader study. Specifically, five images from each case were selected for visual
quality assessment: the original 2-average noisy image, the original 8-average
clinically acquired noisy image, the 2-average denoised image using 1-to-1 repetition training, the
2-average denoised image using 2-to-2 repetition training, and the 2-average
denoised image using 4-to-4 repetition training (the same denoisers from the
previous experiment). This resulted in a total of 100 images evaluated across
20 test subjects. Two experienced abdominal radiologists (H.C., with 17 years
of clinical experience post-fellowship, and L.G., with 3 years) independently
and blindly scored each image based on overall image quality, sharpness of the
prostate capsule, and artifact/noise level. A 4-point Likert scale was used,
where a score of 4 indicated the best quality, highest sharpness, or lowest
artifact/noise level, and a score of 1 indicated the worst. The scores from
both readers were averaged for comparison of different methods in each assessment
category. Statistical differences were assessed using the Wilcoxon signed-rank
test, and inter-reader agreement was evaluated by computing the Cohen’s kappa
coefficient.

\section{Results}\label{sec-results}
\begin{figure*}[tbh]
\centerline{\includegraphics[width=0.9\textwidth]{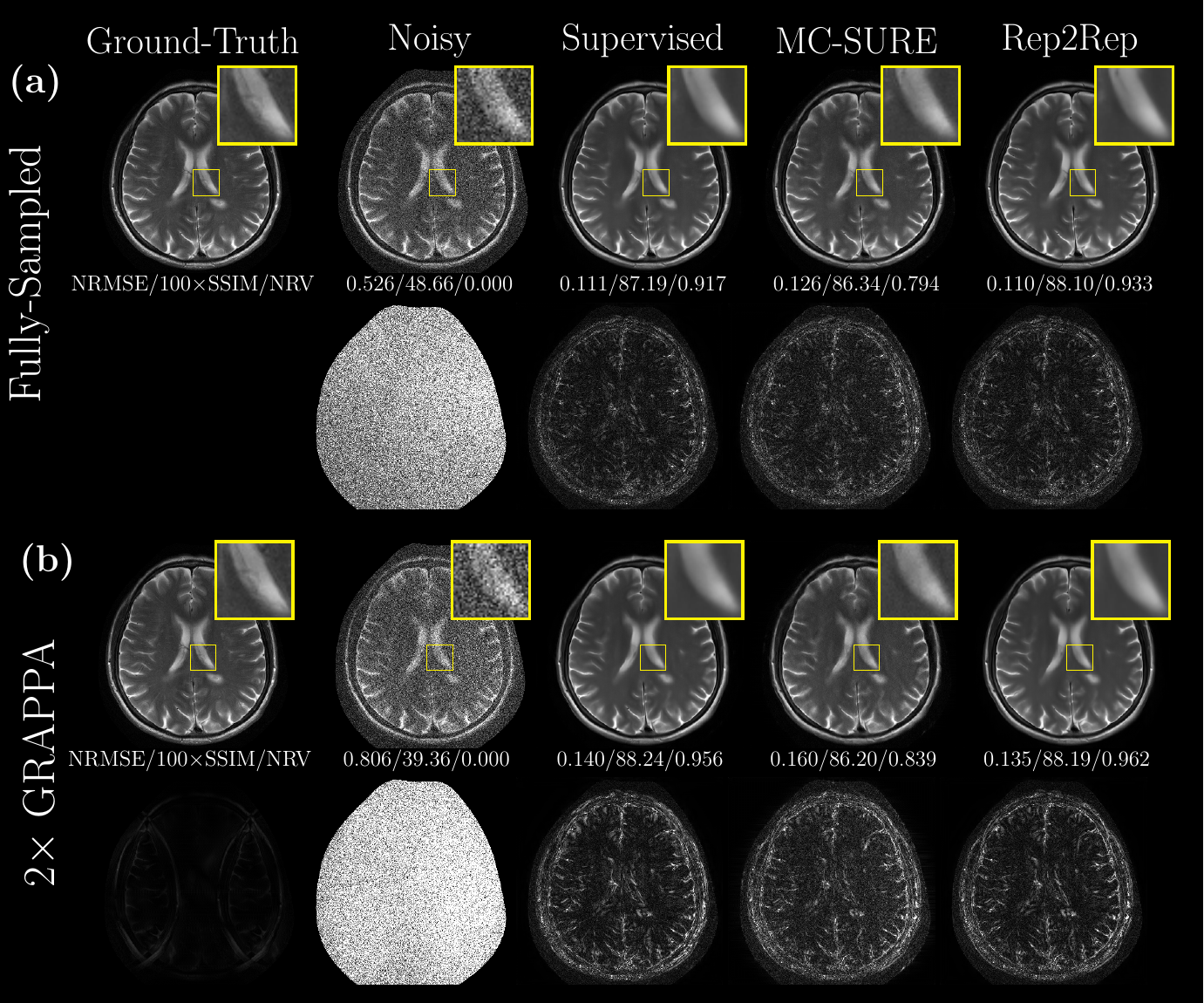}}
\caption{%
    Visual comparison of CDLNet denoising under different training schemes on
    the synthetic-noise T2w fastMRI Brain dataset. (a) Denoising of
    fully-sampled noisy data. (b) Denoising of $2\times$ GRAPPA reconstructed
    data (ACS included). $5\times$ absolute denoising error (compared to the
    noise-free ground-truth) is shown below each image. Quantitative metrics
    (NRMSE/$100\times$SSIM/NRV) are shown below each image. Note that NRMSE and
    SSIM were computed against the ground-truth fully-sampled image and NRV was
    computed against the noisy image, for both fully-sampled and GRAPPA
    denoisings. 
\label{fig-results_syn_grappa}}
\end{figure*}
\begin{figure*}[tbh]
\centerline{\includegraphics[width=\textwidth]{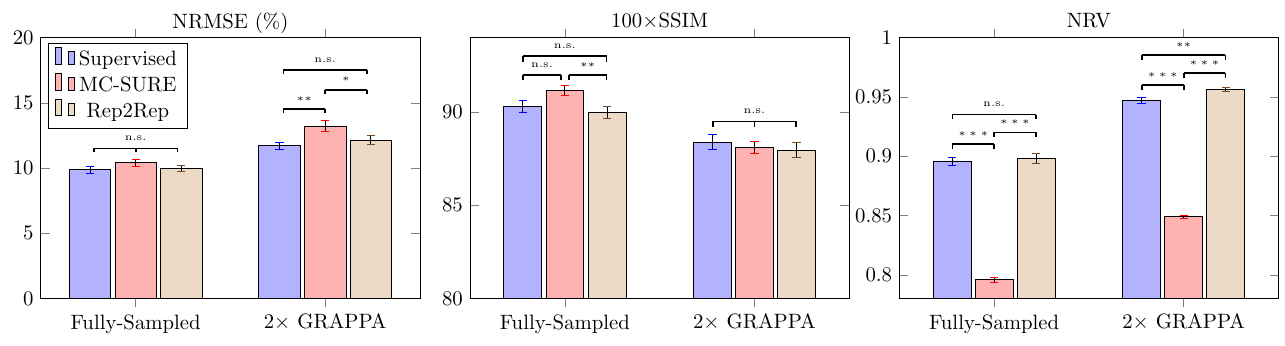}}
\caption{%
Quantitative metrics for synthetic denoising averaged on the fastMRI T2w Brain dataset.
Rep2Rep method trained on 1-repetition noisy inputs and targets.
MC-SURE method trained on 2-avg noisy data.
Supervised method trained on 2-avg noisy inputs with noise-free fully-sampled targets.
Inference performed on 2-avg data for all methods. 
Optimal values for SSIM and NRV are $1$.
No statistically significant difference between metrics denoted by (n.s.). 
Markers $^\ast$, $^{\ast\ast}$, and $^{\ast\ast\ast}$ indicate $p$-values less than $0.05$, $0.01$, and $0.0001$, respectively. 
\label{fig-tablebar}}
\end{figure*}
\begin{figure}
\centerline{\includegraphics[width=\columnwidth]{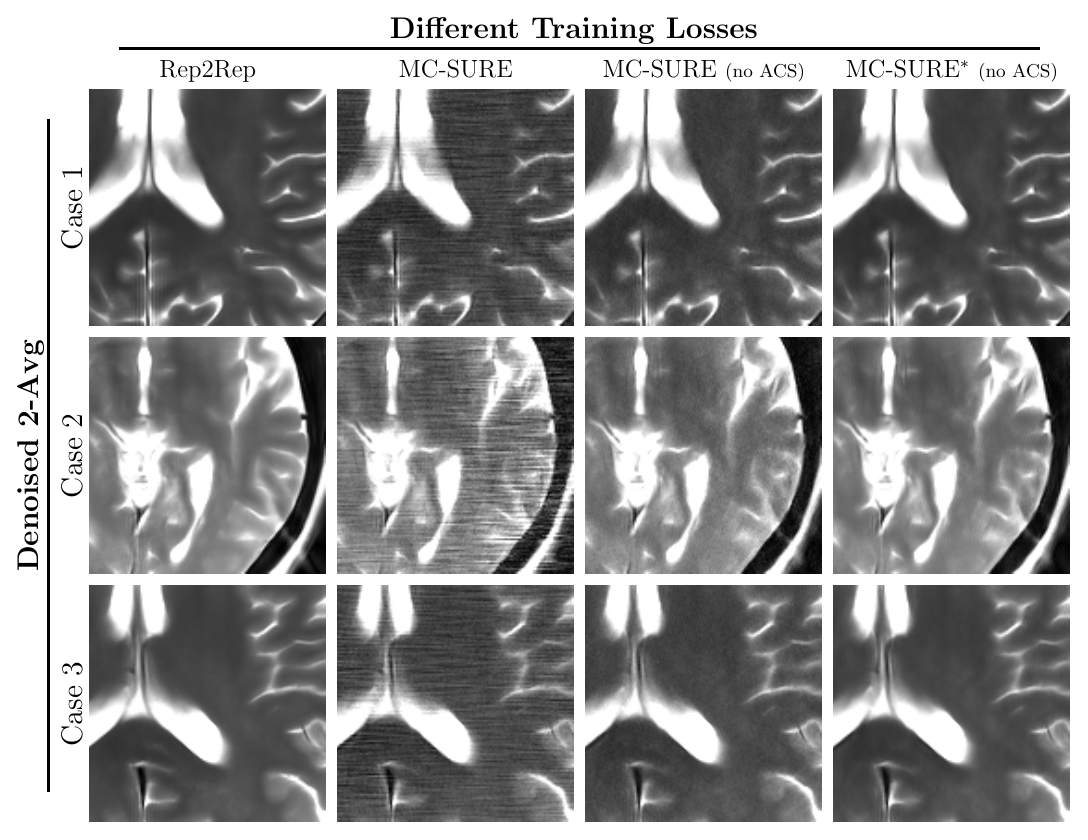}}
\caption{%
    Visual examples on the effect of training loss, inclusion of
    auto-calibration-signal (ACS) and imperfect noise-level information for
    denoising of GRAPPA reconstructed data from the Synthetic-Noise fastMRI T2w
    Brain dataset. All denoisings were performed on GRAPPA+ACS reconstructed data
    using imperfect noise-level information. $^\ast$ indicates that perfect noise-level
    information was used. Inclusion of the ACS region in training data leads to
    streaking artifacts in MC-SURE trained model. With the ACS signal excluded
    from training data (3rd column), denoising appears successful however
    imperfect noise-level information yields a noisy residual artifacts.
    Rep2Rep training handles inclusion of ACS data and imperfect noise-level
    information without issue.
\label{fig-fastmri_fail}}
\end{figure}
\begin{figure*}[tbh]
\centerline{%
    \includegraphics[width=0.56\textwidth]{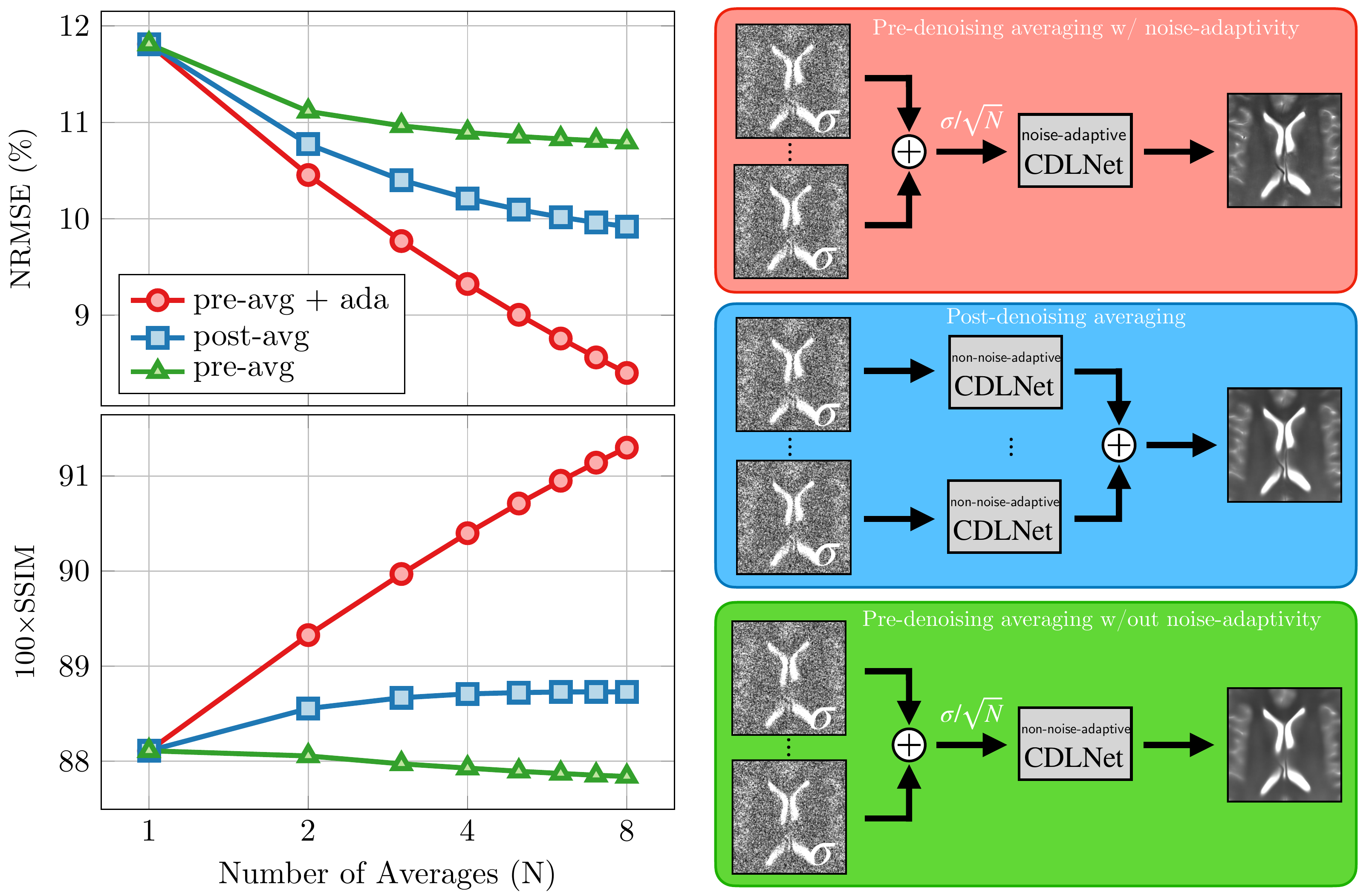}%
    \hfill%
    \includegraphics[width=0.44\textwidth]{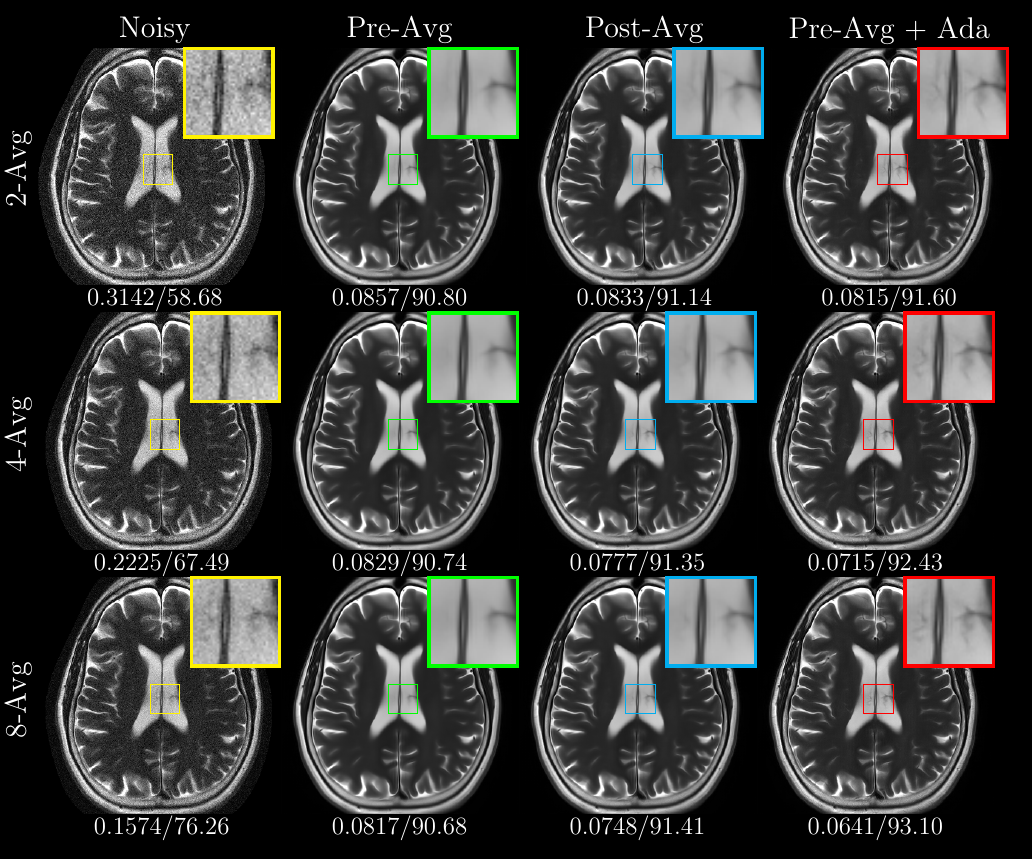}%
}
\caption{%
    Effect of number of repetitions on denoising quality, with different
    denoising+averaging schemes. Left: PSNR (top) / SSIM (bottom) vs. Number of
    Repetitions for a Rep2Rep trained CDLNet model using pre-denoising
    repetition averaging (Pre-Avg+Ada), post-denoising repetition averaging
    (Post-Avg), and pre-denoising repetition averaging without noise-adaptive
    thresholds (Pre-Avg). Middle: graphical legend of the 3 denoising +
    averaging schemes. Right: visual example of the different schemes under 2,
    4, and 8 repetitions available, with NRMSE/100$\times$SSIM shown below each
    image. All methods were trained on fully sampled 1-repetition data from the
    synthetic-noise fastMRI T2w Brain dataset using the Rep2Rep loss.
\label{fig-inference_avg}}
\end{figure*}
\begin{figure}[tbh]
\centerline{\includegraphics[width=\columnwidth]{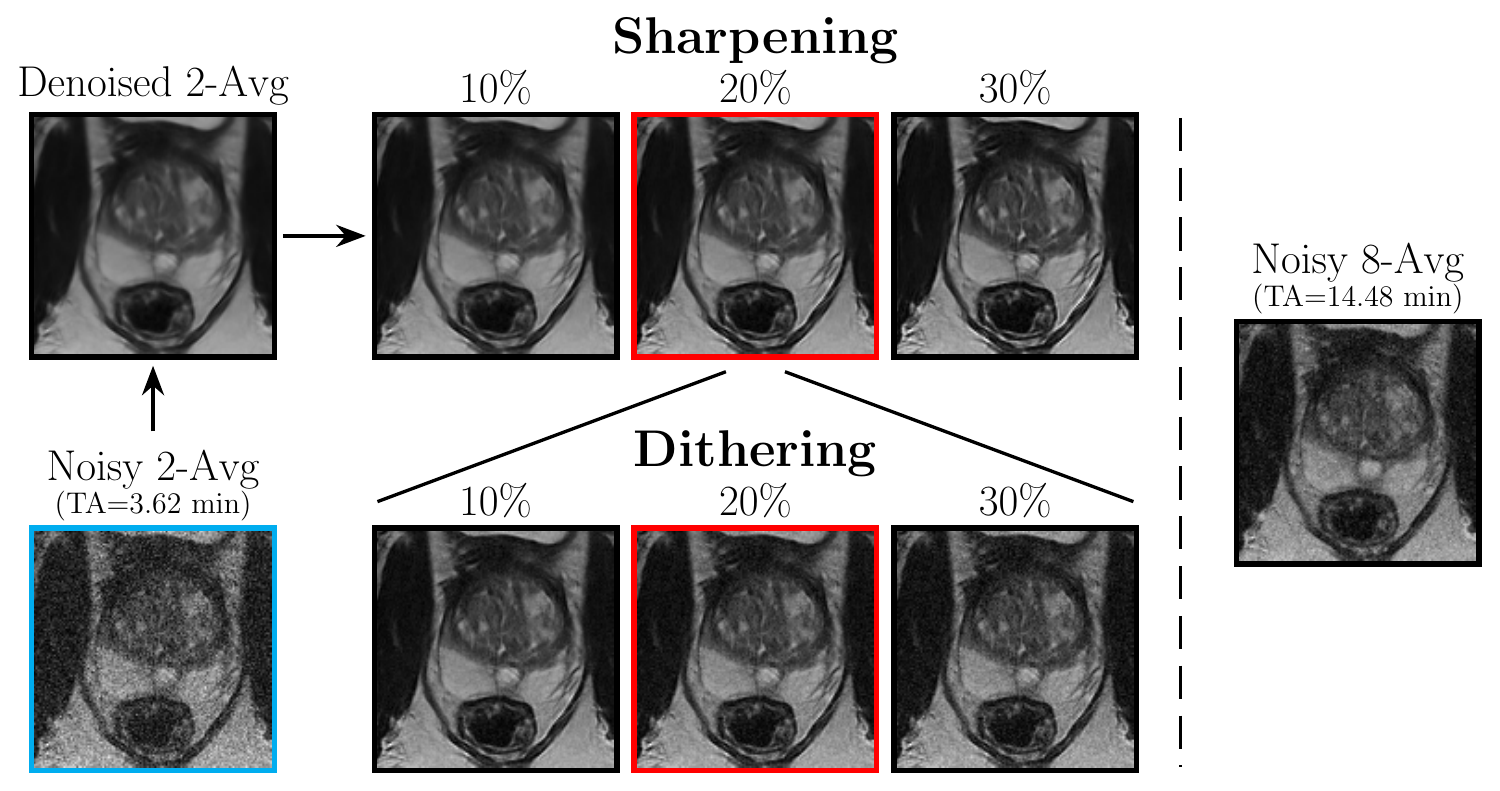}}
\caption{%
    Example of T2w prostate denoising post-processing pipeline. Starting with a
    noisy 2-average image, we denoise, sharpen, and dither. A sharpening of
    $20\%$ and dithering of $20\%$ was selected for all scans in the prostate
    dataset. The noisy 8-average image is shown for reference.
\label{fig-sharp_dither}}
\end{figure}
\begin{figure*}
    \centering
    \begin{subfigure}{0.49\textwidth}
        \centerline{\includegraphics[width=\columnwidth]{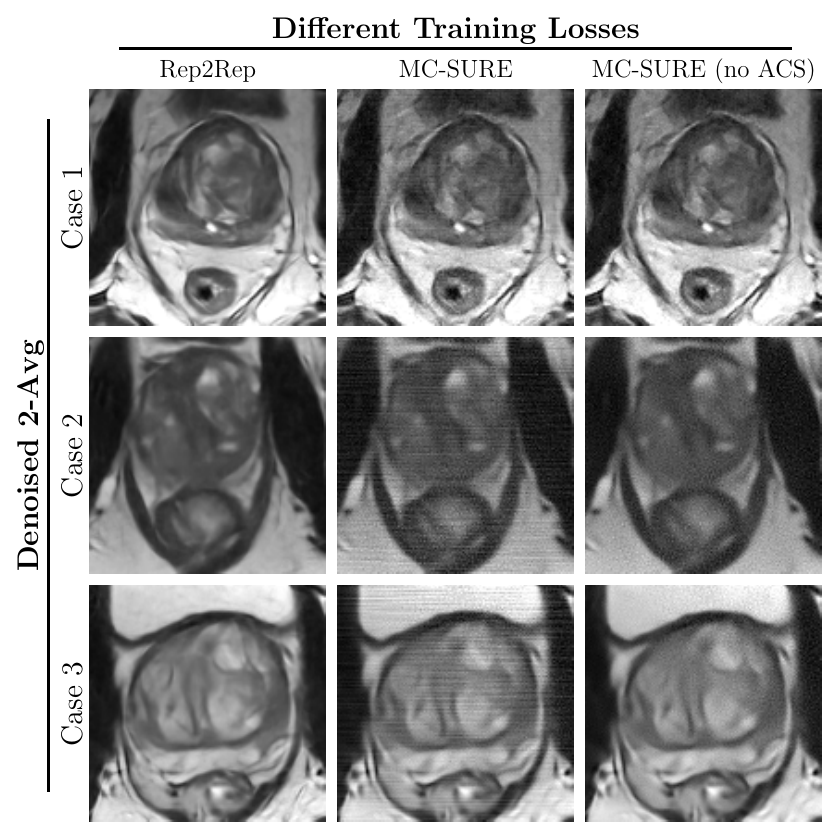}}
        \caption{denoising without post-processing \label{fig-prostate_fail_nodither}}
    \end{subfigure}%
    \hfill%
    \begin{subfigure}{0.49\textwidth}
        \centerline{\includegraphics[width=\columnwidth]{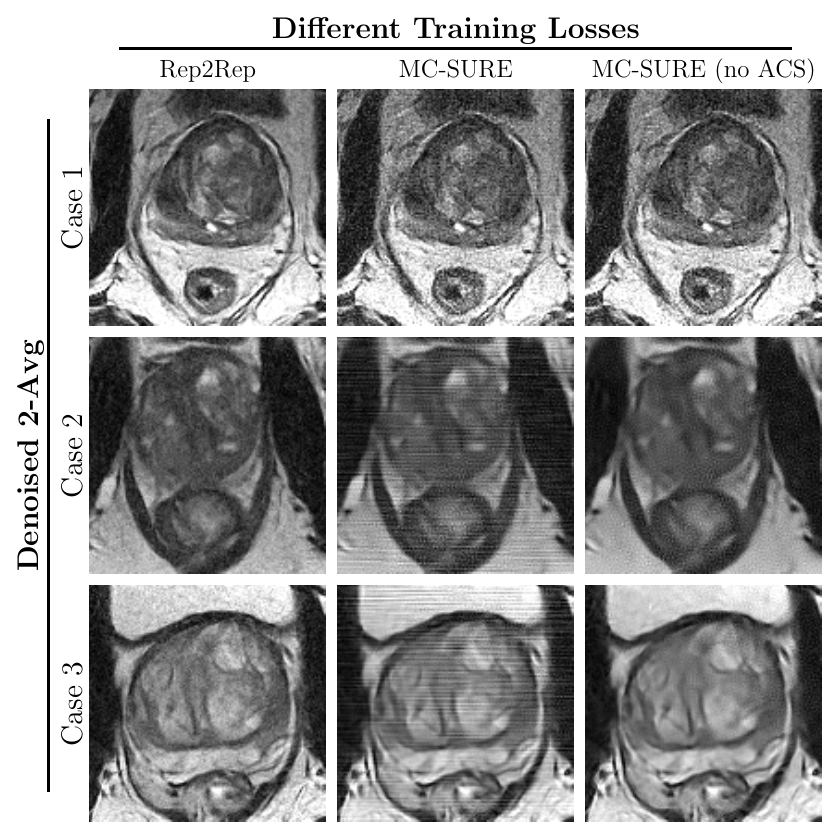}}
        \caption{denoising with post-processing}
        \label{fig-prostate_fail_dither}
    \end{subfigure}%
    \caption{%
        Ablation study corresponding to experimental setup from synthetic-noise
        experiments (Figure~\ref{fig-fastmri_fail}), now using real data from
        0.55T T2w Prostate Dataset. Left: denoising without post-processing.
        Right: denoising with post-processing. Similar qualitative results are
        observed as in the synthetic experiments. Note that this is low-SNR
        data and only imperfect noise-level information (from a reference
        noise-scan) is available. Rep2Rep training sidesteps issues of ACS
        inclusion and imperfect noise-level information, yielding visually
        superior results with and without post-processing.
    }
    \label{fig-prostate_fail}
    \hfill%
\end{figure*}
\begin{figure}[tbh]
    \centering
    \begin{subfigure}{\columnwidth}
        \centerline{%
            \includegraphics[width=\columnwidth]{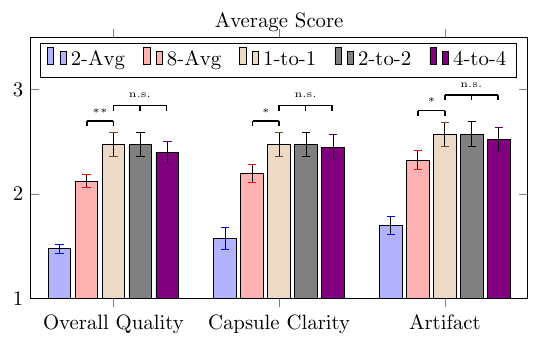}
        }
        \caption{%
            Average radiologist scores with standard error bars shown.
            No statistically significant difference between metrics denoted by (n.s.). 
            Markers $^\ast$ and $^{\ast\ast}$ indicate $p$-values less than $0.05$ and $0.01$, respectively. 
        }
        \label{fig-rstudy_chart}
    \end{subfigure}%
    \hfill%
    \begin{subfigure}{\columnwidth}
        \centerline{\includegraphics[width=\columnwidth]{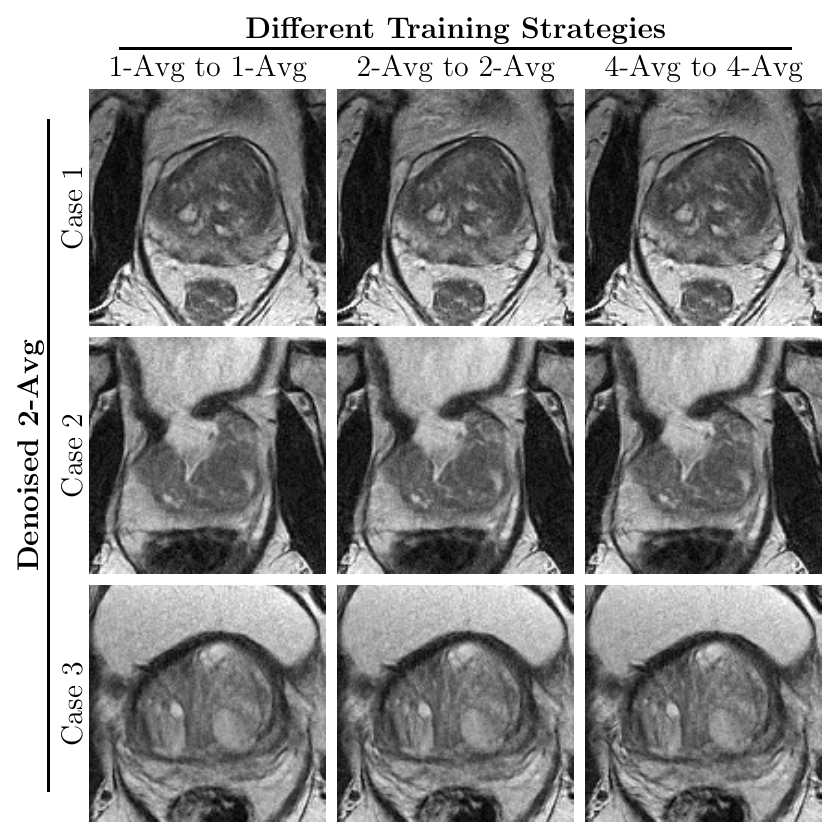}}
        \caption{%
            Denoisings of 2-Avg T2w Prostate data by three Rep2Rep learning models with an
            increasing number repetition data available during training. Left: 1-repetition to
            1-repetition (2 repetitions per scan). Middle: 2-average to 2-average (4 repetitions per
            scan). Right: 4-average to 4-average (8 repetitions per scan). Cases correspond to
            zooms shown in Figure~\ref{fig-prostate_cases}.
        }
        \label{fig-prostate_cases_zoom}
    \end{subfigure}%
    \caption{%
        Radiologist study on 0.55T T2w Prostate testset
        images and denoisings. Left: average score across clinically relevant
        metrics. Right: center-crops from example denoisings used in the study.
    }
    \label{fig-rstudy}
\end{figure}
\begin{figure*}
\centerline{\includegraphics[width=\textwidth]{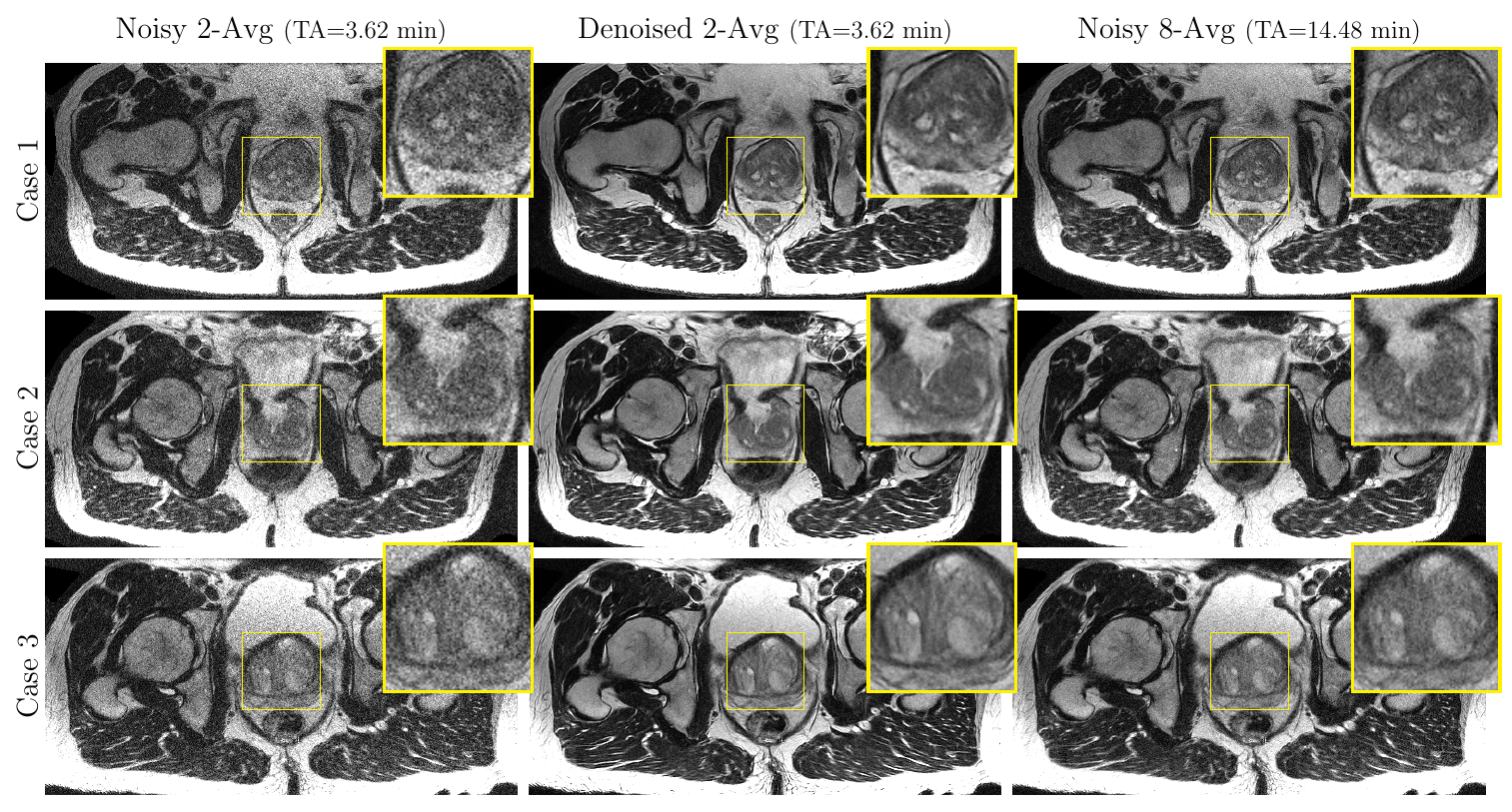}}
\caption{%
    Example denoisings of 2-Avg repetition data from 3 different
    patient scans in the T2w Prostate Image Dataset. Left: Noisy 2-Avg input
    image. Middle: Rep2Rep learning denoisings. Right: Noisy 8-Avg reference image. Total
    scan acquisition time (TA) is given above each set of images.
    Zooms highlight prostate region.
\label{fig-prostate_cases}}
\end{figure*}

\subsection{Experiment 1}
Figure \ref{fig-results_syn_grappa}a  shows an example comparing different training methods on fully
sampled brain MRI with synthetic noise, including the corresponding $5\times$
scaled absolute denoising error maps (i.e. $5\abs{\x - \hat{\x}}$) and
associated quantitative metrics. In this case, Rep2Rep learning achieves
denoising performance comparable to supervised learning. Although MC-SURE
yields similar visual image quality, it results in slightly inferior NRMSE and
SSIM values, along with an ~10\% lower NRV. Results across all testing cases
are summarized in Figure \ref{fig-tablebar} (see the fully sampled bar plots).

\subsection{Experiment 2}
Figure \ref{fig-results_syn_grappa}b shows an example from the same case as in
Figure \ref{fig-results_syn_grappa}a, comparing different training methods on
$2\times$ accelerated brain images reconstructed using GRAPPA. In this case,
Rep2Rep learning maintains performance comparable to supervised learning, while
MC-SURE yields inferior denoising quality, as indicated by the error map and
quantitative metrics. Results across all testing cases are summarized in Figure
\ref{fig-tablebar} (see the $2\times$ GRAPPA bar plots). Similar to the fully
sampled case, MC-SURE results in an ~10\% lower NRV than both supervised and
Rep2Rep learning. 

\subsection{Experiment 3}
\label{sec-results-fastmri-fail}
Figure \ref{fig-fastmri_fail} shows three cases comparing Rep2Rep learning (first column) with
MC-SURE denoising applied to GRAPPA-reconstructed images with and without
inclusion of ACS data (second and third columns), all at $2\times$ acceleration. For
Rep2Rep learning, the ACS data was included in the final reconstruction. The
denoising results in columns 1–3 were generated using estimated noise levels.
These results indicate that MC-SURE leads to reduced denoising performance on
GRAPPA-reconstructed images, primarily due to the introduction of local spatial
noise correlation when ACS data is included, as shown in the Supporting
Information. While spatial noise correlation also exists in images
reconstructed without the ACS region, it is predominantly non-local and thus
has minimal impact on MC-SURE denoising. In contrast, Rep2Rep learning is not
affected by such noise correlation. 

The fourth column of Figure \ref{fig-fastmri_fail} shows the
MC-SURE results when perfect noise-level information is provided (denoted as
MC-SURE$^\ast$), using the same noise level used for synthesis. In this case, MC-SURE
achieves denoising quality comparable to that of Rep2Rep learning. 
In practical applications, perfect noise-level estimation is not achievable, which highlights a
fundamental limitation of MC-SURE denoising.

\subsection{Experiment 4}
Figure \ref{fig-inference_avg} presents the corresponding results
using varying numbers of repetitions during inference, with all models trained
using a 1-to-1 repetition configuration. When inference is performed using a
single repetition, all three methods yield identical results, as the noise
level matches that seen during training. However, as the number of repetitions
increases, the noise level at inference deviates more significantly from that
during training. Among the three approaches, the proposed method (Pre-Avg+Ada) consistently achieves the
best performance across all repetition settings. 

\subsection{Experiment 5}
Figure \ref{fig-sharp_dither} presents results from a representative case to illustrate the effects
of sharpening and dithering at varying strengths. The sharpening step visibly
enhances the delineation of fine structures and mitigates image smoothing. A
sharpening level of 20\% was selected by the radiologist and applied across all
testing cases. Dithering was then applied to the sharpened image by
reintroducing the original noisy image with reduced intensity. This process
leads to more natural-looking results for human perception. A dithering level
of 20\% was also selected by the radiologist for use in all testing cases.

\subsection{Experiment 6}
Figure \ref{fig-prostate_fail_nodither} compares Rep2Rep learning (with ACS data included in the final
images; first column) and MC-SURE denoising with and without ACS data inclusion
(columns 2 and 3, respectively) across three cases. Consistent with findings
from the synthetic brain datasets (see Figure \ref{fig-fastmri_fail}), MC-SURE without ACS inclusion performs
poorly due to imperfect noise-level estimation, while MC-SURE with ACS
inclusion shows further degraded performance due to the introduction of local
spatial noise correlation. Figure \ref{fig-prostate_fail_dither} presents the same comparison after
applying post-processing using the sharpening and dithering settings selected
in Experiment 5. Similar trends are observed, with Rep2Rep learning
consistently outperforming both MC-SURE results.

\subsection{Experiment 7}
Figure \ref{fig-prostate_cases_zoom} shows three cases comparing Rep2Rep training with different numbers of
repetitions: 1-to-1, 2-to-2, and 4-to-4. As described in the Methods section,
all models were evaluated using 2-repetition averaged images during inference.
Across all cases, the denoising quality appears visually comparable among the
three training configurations. These results suggest that the proposed method
is robust to the number of repetitions used during training, and that training
with higher-SNR inputs, such as two-average or four-average images, does not
yield significant improvement in denoising performance.

\subsection{Experiment 8}
Figure \ref{fig-rstudy_chart} presents the qualitative comparison based on the reader study to
assess the original noisy images against Rep2Rep denoised images trained with
different repetition numbers across all testing cases. All Rep2Rep results were
generated using two-repetition averaged images during inference. The results
show that Rep2Rep learning significantly outperforms the original
8-average noisy images across all three assessment categories
(overall $p=0.0093$; capsule $p=0.0278$; artifact $p=0.0444$). 
Moreover, no significant differences were observed
between the Rep2Rep models trained with different repetition counts across all
metrics ($p>0.4$), confirming the findings from Experiment 7. The inter-reader
agreement was 54.7\%, with an expected agreement by chance of 40.8\%, resulting
in a kappa value of 0.2343 (95\% CI: 0.1392 to 0.3295), indicating fair
agreement. The maximum possible kappa value, given the marginal distributions,
was 0.8030, with the observed kappa representing 29.2\% of this maximum. 

Figure \ref{fig-prostate_cases} shows results from three representative cases comparing the original
2-average and 8-average noisy images to Rep2Rep-denoised images (trained with
1-to-1 repetition). In all three cases, Rep2Rep learning yields visibly
superior image quality, further supporting the findings from the reader study
shown in Figure \ref{fig-rstudy_chart}.

\section{Discussion}\label{sec-discussion}
Several key features contribute to the performance of Rep2Rep learning. First,
CDLNet allows for robust denoising and generalization across noise levels. This is important for in our target
applications, because multiple repetitions are separated during training but
averaged during inference to increase baseline SNR. This leads to a discrepancy
in noise levels that requires a model with noise-level extrapolation capabilities. Second, unlike
competing self-supervised denoising methods that require accurate estimation of noise level,
Rep2Rep learning has demonstrated good performance even when noise estimation
is imperfect. Third, Rep2Rep learning is compatible with spatially correlated
noise, which is commonly introduced by parallel imaging techniques such as
GRAPPA. This feature is expected to be crucial for clinical deployment, as
parallel imaging is used routinely in MRI exams. 

An important finding from our study is that MC-SURE-based denoising fails when
applied to GRAPPA-reconstructed images that include the ACS data, which is
typically retained to ensure higher SNR. This is because the inclusion of ACS
regions introduces spatially correlated noise, thus violating the assumption
MC-SURE on independent noise distributions, as demonstrated in Supporting
Information. In addition, MC-SURE relies on accurate noise-level estimation,
which is often difficult to obtain in real-world scenarios. In contrast,
Rep2Rep learning overcomes these challenges by employing the Noise2Noise
paradigm, which is robust to spatial noise correlation and does not require
accurate noise-level estimation, making it more practical for clinical
applications. 

Like other denoising techniques, Rep2Rep may introduce spatial blurring due to
the use of an MSE-based loss function, which tends to smooth image details. To
mitigate this effect, we incorporated two post-processing steps, including
sharpening and dithering, into our denoising pipeline. The dithering step is
performed by reintroducing information from the original image, which helps
restore fine structural details than simply adding artificial noise as
commonly done in deep learning-based image reconstruction. Together, these
post-processing steps improve the overall visual quality of denoised images,
making them more suitable for clinical evaluation.  

Beyond standard denoising, Rep2Rep learning could be potentially extended for
self-supervised joint denoising and reconstruction with repetition
data~\cite{yaman_self-supervised_2020,Spicer_Hu2024}. This could eliminate the
need for an explicit pre-reconstruction step, such as GRAPPA, and may also
enable higher acceleration rates to further reduce scan time, at the cost of a
less flexible post-acquisition pipeline.
Rep2Rep learning could further be applied to quantitative imaging applications
where quantification accuracy is highly sensitive to imaging SNR~\cite{kang_self-supervised_2024, Zhang2022Blip}.
In addition, Rep2Rep learning could be applied to other MRI applications that
rely on multiple signal averages, such as diffusion-weighted imaging (DWI)~\cite{kang_self-supervised_2024,Pfaff2024},
which also suffers from inherent low SNR. Future work will explore these
extensions in greater detail. 

While Rep2Rep learning demonstrates strong potential, several limitations
warrant discussion. First, the method relies on multiple repetitions during
training, restricting its applicability to MRI protocols that involve
multi-repetition acquisitions. A potential extension to address this limitation
is the incorporation of a coil-to-coil loss, as explored in our preliminary
studies. Second, although Rep2Rep is noise-adaptive, its performance may still
depend on the range and distribution of noise levels encountered during
training. Whether the model remains effective under extreme out-of-distribution
noise conditions remains an open question and requires further investigation.
Third, this study focused only on brain and prostate MRI datasets. Further
validation across more anatomies, imaging contrasts, and field strengths is
needed to fully assess the clinical utility of this method.

\section{Conclusion}\label{sec:conclusion}
This work presented Rep2Rep learning, a deep learning-based denoising technique
that leverages the repeated acquisitions commonly performed in low-field MRI,
or other low-SNR settings requiring multi-repetition scans, to train a
noise-adaptive denoiser without requiring ground-truth data. Its noise-adaptive
capability enables the model to generalize effectively across different SNR
regimes, depending on the number of repetitions available at inference. Rep2Rep
learning also supports denoising of GRAPPA-reconstructed data with spatially
correlated noise and demonstrates robustness to imperfect noise-level
estimation. These capabilities make Rep2Rep learning a promising approach for
improving image quality and scan efficiency in low-field MRI.

\section*{Acknowledgments}
This work was supported by the NIH (R01EB030549, R01EB031083, R21EB032917, and
P41EB017183) and was performed under the rubric of the Center for Advanced
Imaging Innovation and Research (CAI2R), an NIBIB National Center for
Biomedical Imaging and Bioengineering. The authors would like to thank Amirhossein
Khalilian-Gourtani, PhD for helpful discussion.

\subsection*{Financial disclosure}
None reported.

\subsection*{Conflict of interest}
The authors declare no potential conflict of interests.

\bibliography{references}%
\vfill\pagebreak

\section*{Supporting information}
\section{Notation}
In this manuscript, we consider 2D multi-channel images as vectors.
Specifically, image $\x \in \C^{NC}$, with $N$ pixels and $C$ channels is formed
by stacking the vectorized 2D channels in a column vector,
\begin{equation} \label{eq:img_vec}
\x = \begin{bmatrix}
        \x_1 \\ \x_2 \\ \vdots \\ \x_C
    \end{bmatrix}.
\end{equation}

Operators which operate identically on the channels of an image or the pixels
of an image can be represented as block matrices. For example, we can express
applying the same matrix $\bA \in \C^{N \times N}$ to every channel as, 
\begin{equation}
    (\bA \otimes \IDMAT_C)\x = 
    \begin{bmatrix}
        \bA &     & & \\
            & \bA & & \\
            &     & \ddots & \\
            &     &        & \bA 
    \end{bmatrix}
    \begin{bmatrix}
        \x_1 \\ \x_2 \\ \vdots \\ \x_C 
    \end{bmatrix} 
    =
    \begin{bmatrix}
        \bA\x_1 \\ \bA\x_2 \\ \vdots \\ \bA\x_C 
    \end{bmatrix},
\end{equation}
and we say that $(\bA \otimes \IDMAT_C)$ is separable over channels or that $\cwise{\bA}$ is a channel-wise operator. Now suppose we
have a matrix $(w_{ij}) = \bW \in \C^{M \times C}$ which operates on image pixels $\x[i]
\in \C^C, ~ 1\leq i \leq N$, i.e. $\y[i] = \bW(\x[i]) \in \C^M$. We can conveniently express this as,
\begin{equation}
    (\IDMAT_N \otimes \bW)\x = 
    \begin{bmatrix}
        w_{11}I_N & w_{12}I_N & \cdots & w_{1C}I_N \\
        w_{21}I_N & w_{22}I_N & \cdots & w_{2C}I_N \\
        \vdots & \vdots & \ddots & \cdots \\
        w_{M1}I_N & w_{M2}I_N & \cdots & w_{MC}I_N \\
    \end{bmatrix}
    \begin{bmatrix}
        \x_1 \\ \x_2 \\ \vdots \\ \x_C 
    \end{bmatrix}.
\end{equation}
We say that $\pwise{\bW}$ is separable over pixels and is a pixel-wise operator.
Note that $(\IDMAT_N \otimes \bW)$ also corresponds to a $1\times 1$ kernel $C$ to $M$
channel convolution operator.

A nice aspect of this Kronecker product notation is the {\it mixed product property}, which states,
\begin{equation}
    (\bA \otimes \bB)(\bC \otimes \bD) = (\bA\bC \otimes \bB\bD)
\end{equation}
for matrices of compliant sizes. Hence, channel-wise and pixel-wise operators commute, i.e. $\pwise{\bA}\cwise{\bB} = \cwise{\bB}\pwise{\bA}$.

\section{Synthetic Noise Generation}
We generated a single positive semi-definite
noise-covariance matrix $\bSigma$ for each volume using the following scheme,
\begin{equation}
    \bL_{ij} \sim
    \begin{cases}
        \N(\sigma_\mathrm{diag}, \, \sigma_\mathrm{jitter}^2), & i=j \\
        \mathcal{U}(-\sigma_{\mathrm{corr}}/C, \, \sigma_{\mathrm{corr}}/C), & \mathrm{else}
    \end{cases}
\end{equation}
where $\bSigma = \bL\bL^H$. 
Here, $\sigma_\mathrm{diag}$ determines the nominal
noise-level of the image, $\sigma_\mathrm{jitter}$ introduces factors which
contribute to a spatially varying noise in the coil-combined image, and
$\sigma_{\mathrm{corr}}$ explicitly controls correlations between coil noise
vectors, which also contribute to spatially noise variations in final image.
We then generated our noisy volume with R repetitions via,
\begin{equation}
\y_{(r)} = \bS\x + \pwise{\bL}\bbb_{(r)}, \quad \bbb_{(r)} \sim \N(0, \, \IDMAT),~\forall~r=1,\dots,R,
\end{equation}
with covariance matrix parameters $\sigma_{\mathrm{diag}} = 0.15$, $\sigma_{\mathrm{jitter}}
= 0.02$, $\sigma_{\mathrm{corr}} = 0.3$.

\section{Noise-Level Estimation}
A reference noise scan
$\hat{\bxi}^{NC}$ may give us N independent observations of multicoil k-space
noise, from which we can compute the k-space coil noise sample covariance
matrix,
\begin{equation}\label{eq-nle}
\hat{\bSigma} = \frac{1}{N-1}\sum_{n=1}^N \hat{\bxi}[n]\hat{\bxi}[n]^H.
\end{equation}
If a dedicated noise scan is unavailable, an approximate noise-scan can be
obtained from the noisy multicoil image via convolution with a 2D orthogonal
wavelet high-high filter $\bg$, i.e.
${\hat{\bxi}_c = \bg \ast (\bF^H\bk_c)}$.

\section{Noise Whitening}\label{sec-whitening}
It is standard to pre-whiten our data to improve optimize SNR upon
coil-combination. To do so, $\bSigma$ may be estimated directly from an
acquisition without a patient in the scanner or indirectly through filtering
methods \cite{Chang2000}. We form a whitening transform as a scaled Hermitian inverse
square-root of the noise covariance matrix,
\begin{equation}\label{eq:white_mat}
\bW = \sigma \isqrtSigma, \quad \sigma = \frac{\maximum \abs{\bk}}{\maximum \abs{\pwise{\isqrtSigma} \bk}}
\end{equation}
where $\sigma$ is used to maintain the original signal value range.
We obtain the whitened kspace via $\bk^\prime = \pwise{\bW}\bk$, 
and whitened normalized coil-sensitivity map operator 
\begin{equation}
\bR = \cwise{\bZ}^{-1}\pwise{\bW}\bS
\end{equation}
where $\bZ = \sigma\sqrt{\bS^H \pwise{\bSigma}^{-1}\bS}$ . Note that $\bZ$ is diagonal and non-zero where $\x$ has
data (i.e. $\bM_x$)\footnote{This allows us to perform division sensibly by
defining $0/0 \equiv 0$}. With this normalization, we perform
coil-combination of the whitened image-domain data and compensate with the sensitivity profile,
\begin{align}
     \y &= \bZ^{-1}\bR^H \cwise{\bF}^H \bk^\prime \\
        &= \bZ^{-1}( \cwise{\bZ}^{-1}\pwise{\bW}\bS )^H \cwise{\bF}^H \pwise{\bW}\bk \\
        &= \bZ^{-1} \bS^H \pwise{\bW}^H \cwise{\bZ}^{-1} \cwise{\bF}^H \pwise{\bW} \bk \\
        &= \bZ^{-2} \bS^H (\bF^H \otimes \bW^H \bW) \bk \\
        &= \bZ^{-2} \bS^H (\bF^H \otimes \sigma^2\bSigma^{-1}) (\cwise{\bF}\bS\x + \bxi) \\
        &= \sigma^2 \bZ^{-2} \bS^H \pwise{\bSigma}^{-1} (\bS\x + \bxi) \\
        &=  (\bS^H\pwise{\bSigma}^{-1}\bS)^{-1}(\bS^H\pwise{\bSigma}^{-1}\bS \x + \bS^H\pwise{\bSigma}^{-1} \bxi)  \\
        &= \x + (\bS^H\pwise{\bSigma}^{-1}\bS)^{-1} \bS^H\pwise{\bSigma}^{-1} \bxi \\
        &= \x + \bxi^\prime_w,
\end{align}
where $\Cov(\bxi^\prime_w) = (\bS^H \pwise{\bSigma}^{-1}\bS)^{-1} = \sigma^2\bZ^{-2}$.
Keeping track of the sensitivity compensation $\bZ$ enables
us to evaluate quantitative metrics (e.g. PSNR, SSIM) with respect to our ground-truth data.

\section{On Correctness of Noise-Level Computations}\label{sec:appendix:correctness}
Throughout this manuscript, we refer to the ``noise-level'' of coil-combined MRI
signals. In general, an image contaminated with Gaussian-noise may have an
arbitrary noise-covariance matrix associated with it, where non-zeros off the
matrix's diagonal would express correlations between noise in different spatial
locations in the image. We may still refer to noise-level
as the square-root of the diagonal of the covariance matrix, which expresses
the noise-power in each spatial location regardless of spatial correlations.

In this section we provide derivations for the noise-level of coil-combined
image-domain MRI. The correctness of these formulas was verified empirically via
simulation. In simulation, an empirical noise-level map may be obtained as
follows. Consider an image $\bmu$ which we contaminate with zero-mean Gaussian
resulting in noisy-image $\y \sim \N(\bmu, \bSigma)$. This image $\y$
may be obtained in some complicated way, such as contaminating $\bmu$ with
noise in kspace followed by inverse Fourier-transform and coil-combination, and
even involving some kspace interpolation. If we generate $R$ such realizations
of these images with noise following the same distribution,
$\{\y_{(r)}\}_{r=1}^R$, we may then obtain an empirical noise-level map via
computing a sample standard-deviation pixel-wise over the repetition dimension, 
\begin{equation}
    \bsigma_{\mathrm{emp}} = \sqrt{\frac{\sum_{r=1}^R \abs{\y_{(r)} - \bmu}^2}{R-1}}.
\end{equation}
As $R$ increases, this empirical noise-level will tend to the true noise-level.
In the following sections, we present derviations for our noise-level formulas
presented in the theory section of the manuscript.

\subsection{Fully-Sampled Data}
For the fully sampled case, we have
\begin{equation} 
\y = \bS^H\cwise{\bF}^H\bk,
\end{equation}
where $\bk \sim \N(\bmu, \, \pwise{\bSigma})$. Hence, 
we have,
\begin{equation}
    \y \sim \N(\bS^H\cwise{\bF}^H\bmu,\, \bS^H\cwise{\bF}^H \pwise{\bSigma} \cwise{\bF}\bS)
    = \N(\x,\, \bS^H\pwise{\bSigma}\bS),
\end{equation}
where the simplification is a direct result of the mixed-product property of the Kronecker product and the normalization of sensitivity map operator $\bS$.

Claim: $\bS^H\pwise{\bSigma}\bS$ is a diagonal matrix. We can show this by seeing that integer translates of the impulse signal, $\bee_i$, form the eigen basis of this matrix, for all $i$:
\begin{align*}
    (\bS^H\pwise{\bSigma}\bS)\bee_i &= (\bS^H\pwise{\bSigma})(\bS\bee_i) \\
        &= \bS^H\pwise{\bSigma}(\bs[i] \circ \bee_i) \\
        &= \bS^H((\bSigma\bs[i]) \circ \bee_i) \\
        &= (\bs[i]^H\bSigma\bs[i]) \bee_i,
\end{align*}
hence the noise covariance matrix is diagonal and given by $\bS^H\pwise{\bSigma}\bS = \diag(\bS^H\pwise{\bSigma}\bs)$.

\subsection{GRAPPA-Reconstructed Data}
\newcommand{\Gconv}{\bG_{\mathrm{conv}}}
\newcommand{\Mconv}{\bM^{\mathrm{conv}}_\Omega}
In the case of subsampled kspace data reconstructed with linear-operator $\bG^H$ (e.g. GRAPPA), we have
\begin{equation} 
\y = \bS^H\cwise{\bF}^H\bG^H\bk,
\end{equation}
where $\bk \sim \N(\bmu, \, \pwise{\bSigma})$ and $\bmu \in \C^{N_sC}$ is the subsampled ground-truth kspace signal ($N_s < N$). 
Hence we have, 
\begin{equation} \label{eq:grappa_sim}
    \y \sim \N(\bS^H\cwise{\bF}^H \bG^H\bmu,\, \bS^H\cwise{\bF}^H\bG^H  \pwise{\bSigma} \bG\cwise{\bF}\bS).
\end{equation}
It is not immediately clear from \eqref{eq:grappa_sim} what the noise-level of
this reconstruction is, or what the structure of the noise-covariance matrix is
in general.

To shed insight, we begin by describing the specific subsampling pattern of
kspace data in terms of index-set $\Omega$, where indices in $\Omega$ indicate
measured kspace locations, labeling the subsampling operator (a row-removed
identity matrix) as $\IDMAT_\Omega$. In the case of GRAPPA (without ACS
inclusion), the interpolation operator may be written as a convolution
following a zero-filling operator, i.e. $\bG^H =
\Gconv^H\cwise{\IDMAT}^T_\Omega$. By convolution theorem, $\Gconv^H$ is
diagonalizable by a channel-wise Fourier transform, $\Gconv^H =
\cwise{\bF}\bLambda\cwise{\bF}^H$, where $\bLambda$ is an operator that
operates independently on pixels in the multi-channel image domain, not mixing
information between pixel locations, i.e. $(\bLambda\x)[i] = \bLambda_i(\x[i])$,
$\bLambda_i \in \C^{C\times C}$, for $1 \leq i \leq N$.

We can now describe the noise-covariance matrix of the coil-combined reconstructed image as,
\begin{align}
    \Cov(\y) &= \bS^H\cwise{\bF}^H\bG^H  \pwise{\bSigma} \bG\cwise{\bF}\bS \\
    &= \bS^H\cwise{\bF}^H\Gconv^H\cwise{\IDMAT}_\Omega^T  \pwise{\bSigma} \cwise{\IDMAT}_\Omega\Gconv\cwise{\bF}\bS \\
    &= \bS^H\cwise{\bF}^H\Gconv^H (\bM_\Omega \otimes \bSigma) \Gconv\cwise{\bF}\bS \\
    &= \bS^H\bLambda^H (\bF^H\bM_\Omega\bF \otimes \bSigma) \bLambda\bS \\
    &= \bS^H\bLambda^H (\Mconv \otimes \bSigma) \bLambda\bS \label{eq-cov-grappa}
\end{align}
where $\Mconv = \bF^H\bM_\Omega\bF$ is the channel-wise image-domain
convolution operator defined by the kspace subsampling mask. This simplified
formula for the covariance matrix makes it clear that the structure is
\textit{\bf not diagonal}. The image-domain convolution operator will have $A$ evenly spaced delta's corresponding to the uniform acceleration rate, with each delta scaled by $1/A$. The diagonal of the covariance matrix will thus be given by, 
\begin{equation}
    \bsigma = \sqrt{\frac{1}{A}\bS^H\bLambda^H \pwise{\bSigma} \bLambda \bs}.
\end{equation}
This expression can be obtained by considering passing the $i$-th column of the
identity matrix through the full covariance operator, and only keeping the
$i$-th entry. 

GRAPPA is most commonly used in conjunction with auto-calibration signal (ACS) at the center of kspace. This ACS region is used as reference data to determine the weights of the GRAPPA interpolation kernel, and it is sensible to include this acquired signal as part of the final reconstruction. Let $\Omega$ denote the index set of all acquired samples. We decompose this index set into a $A\times$ uniform subsampling mask $\Xi$ and an ACS region index set $\Theta$ which are not disjoint, i.e. $\Omega = \Xi \cup \Theta$ and $\Xi \cap \Theta \neq \emptyset$. 
We can write the interpolation operator as, $\bG^H = (\Gconv^H\cwise{\bM}_{\bar{\Theta}} + \cwise{\bM}_\Theta) \cwise{\IDMAT}^T_\Omega$, where $\bar{\Theta}$ denotes the complement of set $\Theta$. The resulting noise covariance matrix from this GRAPPA+ACS reconstruction can be described as,
\begin{align}
    &~ \Cov(\bG^H\bk) \\
    &= \bG^H  \pwise{\bSigma} \bG \\
    &= (\Gconv^H \cwise{\bM}_{\bar{\Theta}} + \cwise{\bM}_\Theta)
        (\bM_\Omega \otimes \bSigma) 
        (\cwise{\bM}_{\bar{\Theta}}\Gconv + \cwise{\bM}_\Theta) \\
    &= 
        \Gconv^H (\bM_{\Omega\cap\bar{\Theta}} \otimes \bSigma) \Gconv + (\bM_\Theta \otimes \bSigma).
\end{align}
Hence the image-domain coil-combined noise covariance will be described by,
\begin{align}
    &~ \Cov(\y) \nonumber \\
    &= \bS^H(\bLambda^H(\bM^{\mathrm{conv}}_{\Omega\cap\bar{\Theta}} \otimes \bSigma)\bLambda + (\bM^{\mathrm{conv}}_{\Theta} \otimes \bSigma))\bS. \label{eq-cov-grappa-acs}
\end{align}
Clearly this matrix is not diagonal due to the presence of the image domain
convolution operators $\bM^{\mathrm{conv}}_{\Omega\cap\bar{\Theta}}$ and
$\bM^{\mathrm{conv}}_{\Theta}$. However, we may reason about the diagonal of
this covariance matrix by considering applying it to the $i$-th column of the
identity. The convolution operators are our primary interest, as the operators
$\bS$, $\bLambda$, and $\bSigma$ are all pixel-wise (though perhaps spatially
varying) operators, and thus do not contribute to introducing correlations.
First, $\bM^{\mathrm{conv}}_\Theta$ is the convolution operator corresponding
to a low-pass filter in the phase-encoding direction with factional-width $f =
N_{\mathrm{ACS}} / N_2$ (where $N_2$ is the width of kspace in the
phase-encoding direction). Hence, $\bM^{\mathrm{conv}}_\Theta$ is a sinc filter
with peak amplitude $p$. Second, $\bM^{\mathrm{conv}}_{\Omega \cap
\bar{\Theta}}$ can be rewritten as the filter operator corresponding to the
kspace mask $\bM_{\Omega \cap \bar{\Theta}} = \bM_\Xi ( \IDMAT - \bM_\Theta )$,
i.e. the $A\times$ aliasing filter convolved with the impulse filter minus the
$p$-amplitude sinc filter. This combined filter will have an amplitude of
$(1-p)/2$ at its center. All together, thinking only about the $i$-th element
of the result multiplication of the entire covariance matrix with the $i$-th
column of the identity, we can arrive at the following image-domain
coil-combined noise-level formula for an $A\times$ GRAPPA+ACS reconstruction,
\begin{equation}
    \bsigma =\sqrt{\tfrac{1-p}{A} \cdot \bS^H\bLambda^H \pwise{\bSigma} \bLambda \bs + p \cdot \bS^H \pwise{\bSigma} \bs }.
\end{equation}
Again, we emphasize that this noise-level is simply the square-root of the diagonal of a non-diagonal noise-covariance matrix. 

\begin{figure}[tbh]
\centerline{\includegraphics[width=1.1\columnwidth]{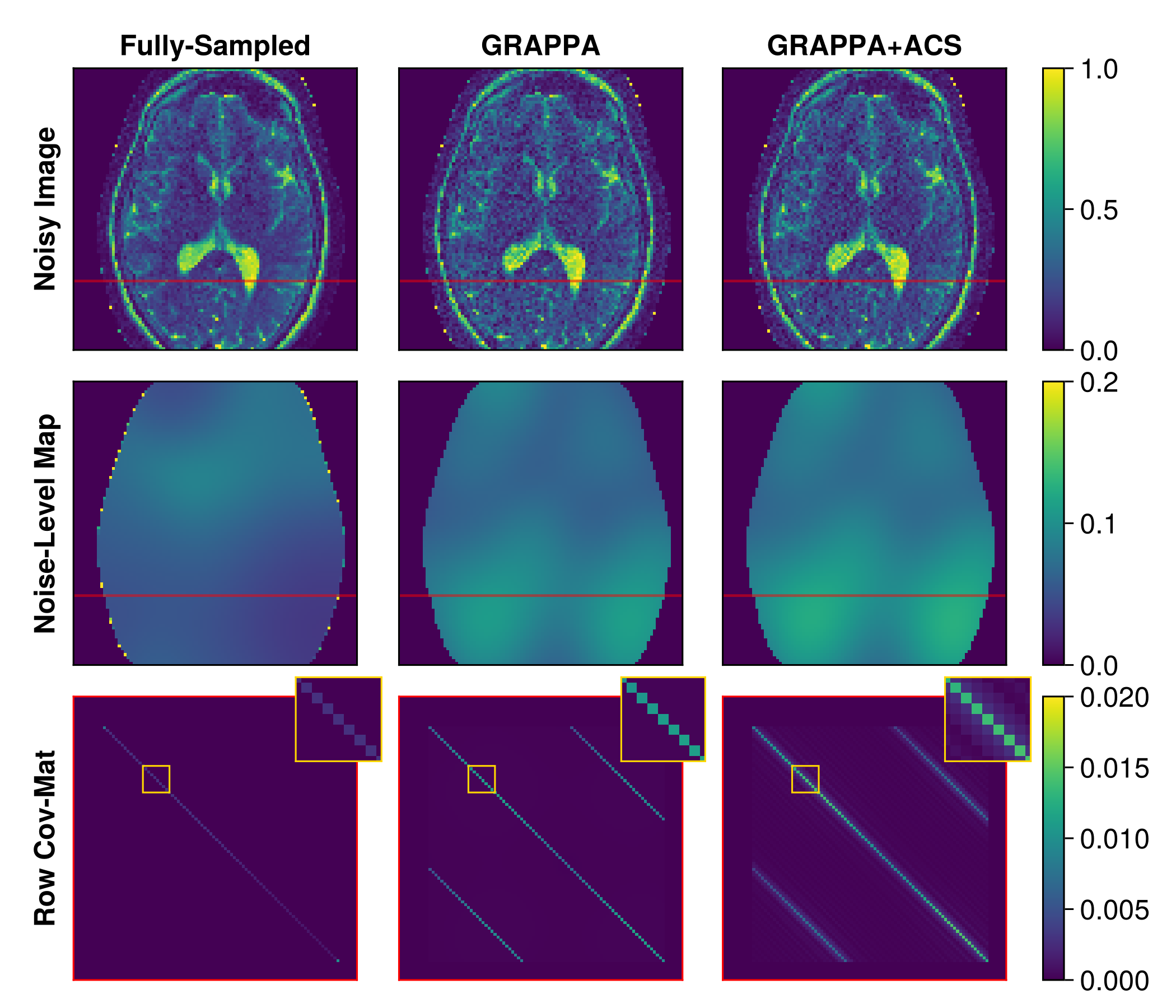}}
\caption{%
    Noise covariance under fully-sampled, GRAPPA reconstructed, and GRAPPA+ACS
    reconstructed settings. Row 1: noisy images. Row 2: associated noise-level
    maps (square-root of noise covariance diagonal). Row 3: sub-matrix of
    noise-covariance matrix corresponding to a single row (highlighted in red)
    of the image. Zoomed in region on the covariance matrix highlights the
    existence of {\it local noise-correlations} in the GRAPPA+ACS reconstructed
    data, whereas fully-sampled and GRAPPA reconstructed data have zero
    off-diagonal elements or only nonlocal correlations, respectively.
    Image colorbars for are given on the right side for each row.
\label{fig-cov_theory}}
\end{figure}

Figure \ref{fig-cov_theory} shows an example set of noisy images from
fully-sampled, GRAPPA reconstructed, and GRAPPA+ACS reconstructed data, along
with their associated noise-level maps (row 2), and a subset of the
noise-covariance matrix between pixels of a single image row. To compute the
covariance matrix, we passed columns of the identity matrix through the full
covariance matrix expressions, given in Equations \eqref{eq-cov-grappa} and
\eqref{eq-cov-grappa-acs}. For computational reasons, we scaled our
ground-truth data and synthetic-noise generation to an image-size of
$96\times 96$. Note that for Cartesian subsampling based GRAPPA reconstruction,
noise correlations are introduced by subsampling in the phase-encoding
direction of k-space, therefore noise-correlations only exist in image-space in
that direction (in this case, only between pixels of a given row).

\end{document}